\documentclass[runningheads]{llncs}

\usepackage{mathtools,stmaryrd,amssymb,amsfonts,nicefrac}
\usepackage{bm}
\usepackage{subcaption}
\usepackage{fancybox,framed}
\usepackage{hyperref}
\usepackage[capitalise,noabbrev]{cleveref}
\usepackage{titletoc}
\usepackage{proof}
\usepackage{enumitem}
\usepackage{subfiles}
\usepackage{diagbox}
\usepackage{nicematrix}

\hypersetup{
    colorlinks,
    linkcolor={blue!80!black},
    citecolor={blue!80!black},
    urlcolor={blue!80!black}
}
\setlist[itemize]{noitemsep, topsep=-.6em}
\allowdisplaybreaks

\usepackage{tikzit}

\tikzstyle{gate}=[fill=white, draw=black, shape=rectangle, minimum height=0.43cm, minimum width=0.43cm, inner sep=0.1em]
\tikzstyle{control}=[fill=black, draw=black, shape=circle, scale=0.38]
\tikzstyle{not}=[shape=circle, path picture={ 
\draw[black](path picture bounding box.north) -- (path picture bounding box.south) (path picture bounding box.west) -- (path picture bounding box.east);
}, draw=black, scale=.8]
\tikzstyle{wcontrol}=[fill=white, draw=black, shape=circle, scale=0.38]
\tikzstyle{bwcontrol}=[draw=black, shape=circle, scale=0.38, path picture={
        \fill[white] (path picture bounding box.center) circle(0.5);
        \fill[black] (path picture bounding box.center) [radius=0.5] -- ++(0:0.5) arc[start angle=0, end angle=180, radius=0.5] -- cycle;
    }, rotate=45]
\tikzstyle{empty}=[fill=white, draw=black, shape=rectangle, inner sep=0.4em, emptyborder]
\tikzstyle{globalphase}=[fill=white, draw=black, inner sep=0.15em, shape=rounded rectangle, minimum height=0.4cm]
\tikzstyle{ancilla}=[fill=black, draw=black, shape=rectangle, minimum width=0.01cm, minimum height=0.25cm, inner sep=0.01em]
\tikzstyle{ground}=[fill=white, path picture={\draw[black](-1.5mm,0)--(-0.6mm,0);\draw[black,thick](-0.6mm,-1.75mm)--(-0.6mm,1.75mm) (0mm,-0.9mm)--(0mm,0.9mm) (0.6mm,-0.5mm)--(0.6mm,0.5mm);}, minimum width=0.1mm, draw=none, outer sep=0pt]
\tikzstyle{gate22}=[fill={rgb,255: red,220; green,220; blue,220}, draw=black, shape=rectangle, minimum height=.68cm, minimum width=0.6cm]
\tikzstyle{void}=[shape=rectangle, minimum height=0.5cm]
\tikzstyle{gate44}=[fill={rgb,255: red,220; green,220; blue,220}, draw=black, shape=rectangle, minimum height=1.43cm, minimum width=0.5cm]
\tikzstyle{divider}=[fill={rgb,255: red,220; green,220; blue,220}, draw=black, shape border rotate=90, regular polygon, regular polygon sides=3, inner sep=1.5pt, rounded corners=0.5mm]
\tikzstyle{gatherer}=[fill={rgb,255: red,220; green,220; blue,220}, draw=black, shape border rotate=-90, regular polygon, regular polygon sides=3, inner sep=1.5pt, rounded corners=0.5mm]
\tikzstyle{hyperedge}=[fill=white, draw=black, shape=rectangle, rounded corners=0.1cm, minimum height=.6cm, minimum width=.6cm]
\tikzstyle{square}=[fill=white, draw=black, shape=rectangle, minimum height=0.20cm, minimum width=0.20cm, inner sep=0.1em, thick]
\tikzstyle{gphase}=[rounded rectangle, rounded rectangle arc length=120, fill={zx_grey}, inner sep=2pt, font={\tiny\boldmath}, label distance=1mm, fill opacity=.8, text opacity=1, tikzit category=ZX]
\tikzstyle{customcontrol}=[fill=white, draw=black, inner sep=0.1em, shape=rounded rectangle, minimum height=0.2cm]
\tikzstyle{whiteancilla}=[fill=white, draw=black, shape border rotate=-90, regular polygon, regular polygon sides=3, inner sep=1.5pt, rounded corners=0.2mm]
\tikzstyle{blackancilla}=[fill=black, draw=black, shape border rotate=-90, regular polygon, regular polygon sides=3, inner sep=1.5pt, rounded corners=0.2mm]
\tikzstyle{greyancilla}=[fill={rgb,255: red,150; green,150; blue,150}, draw=black, shape border rotate=-90, regular polygon, regular polygon sides=3, inner sep=1.5pt, rounded corners=0.2mm]
\tikzstyle{whiteancillaterm}=[fill=white, draw=black, shape border rotate=90, regular polygon, regular polygon sides=3, inner sep=1.5pt, rounded corners=0.2mm]
\tikzstyle{blackancillaterm}=[fill=black, draw=black, shape border rotate=90, regular polygon, regular polygon sides=3, inner sep=1.5pt, rounded corners=0.2mm]

\tikzstyle{emptyborder}=[-, dash pattern=on 0.16em off 0.16em on 0.16em off 0.16em on 0.16em off 0em]
\tikzstyle{etc}=[-, draw=black, densely dashed, thick]
\tikzstyle{greyetc}=[-, draw={rgb,255: red,161; green,161; blue,161}, densely dashed, thick]
\tikzstyle{dots}=[-, dotted, draw=black, thick]
\tikzstyle{big}=[-, thick]
\tikzstyle{register}=[-, double]
\tikzstyle{grey}=[-, draw={rgb,255: red,161; green,161; blue,161}]
\tikzstyle{border}=[-, fill=white]

\input{qc.tikzdefs}
\newcommand{\tf}[1]{\scalebox{0.77}{\input{#1.tikz}}}
\newcommand{\bigtf}[1]{\scalebox{0.90}{\input{#1.tikz}}}
\newcommand{\smalltf}[1]{\scalebox{0.72}{\input{#1.tikz}}}
\newcommand{\tinytf}[1]{\scalebox{0.64}{\input{#1.tikz}}}
\usetikzlibrary{arrows.meta} 
\tikzstyle{inclusion} = [-{Hooks[scale=1.3]}, shorten >=2pt]

\usepackage{xcolor}
\definecolor{darkblue}{rgb}{0.0, 0.0, 0.75}

\newcommand{\N}{\mathbb{N}}

\newcommand{\R}{\mathbb{R}}
\newcommand{\C}{\mathbb{C}}
\newcommand{\ket}[1]{\left| #1 \right\rangle}
\newcommand{\bra}[1]{\left\langle #1 \right|}
\newcommand{\ketbra}[2]{\left| #1 \right\rangle\!\left\langle #2 \right|}

\newcommand{\defeq}{\coloneqq}
\newcommand{\interp}[1]{\llbracket #1 \rrbracket}
\newcommand{\biginterp}[1]{\left\llbracket #1 \right\rrbracket}
\newcommand{\eqcref}[1]{\overset{\cref{#1}}{=}}
\newcommand{\eqoneref}[1]{\overset{\eqref{#1}}{=}}
\newcommand{\eqtworef}[2]{\overset{\eqref{#1}\eqref{#2}}{=}}
\newcommand{\eqthreeref}[3]{\overset{\eqref{#1}\eqref{#2}\eqref{#3}}{=}}
\newcommand{\eqfourref}[4]{\overset{\eqref{#1}\eqref{#2}\eqref{#3}\eqref{#4}}{=}}
\newcommand{\cat}[1]{\textbf{\textup{#1}}}
\newcommand{\et}[1]{\mathcal{#1}}
\newcommand{\ctrl}{\textup{\textsf{C}}}
\newcommand{\invers}[1]{{#1}^{-1}}
\newcommand{\suchthat}{\;\text{s.t.}\;}
\newcommand{\natset}[1]{\{1,\dots,#1\}}
\newcommand{\idmat}[1]{I_{#1}}
\renewcommand{\sin}{\mathsf{sin}}
\renewcommand{\cos}{\mathsf{cos}}
\renewcommand{\arg}{\mathsf{arg}}

\renewcommand{\max}{\mathsf{max}}
\renewcommand{\min}{\mathsf{min}}

\newcommand{\diag}{\mathsf{diag}}

\title{Diagrammatic Reasoning with Control as a Constructor, Applications to Quantum Circuits}
\titlerunning{Diagrammatic Reasoning with Control as a Constructor}
\author{Noé Delorme \and Simon Perdrix}
\authorrunning{N. Delorme and S. Perdrix}
\institute{Université de Lorraine, CNRS, INRIA, LORIA, 54000 Nancy, France}

\begin{document}

\maketitle

\begin{abstract}
    Control is a fundamental concept in quantum and reversible computational models. It enables the conditional application of a transformation to a system, depending on the state of another system. We introduce a general framework for diagrammatic reasoning featuring control as a constructor. To this end, we provide an elementary axiomatization of control functors, extending the standard formalism of PROPs to controlled PROPs. As an application, we show that controlled PROPs facilitate diagrammatic reasoning for quantum circuits by introducing a simple and complete set of relations involving at most three qubits, whereas in the standard PROP setting any complete axiomatization necessarily requires relations acting on arbitrarily many qubits.
\end{abstract}


\section{Introduction}

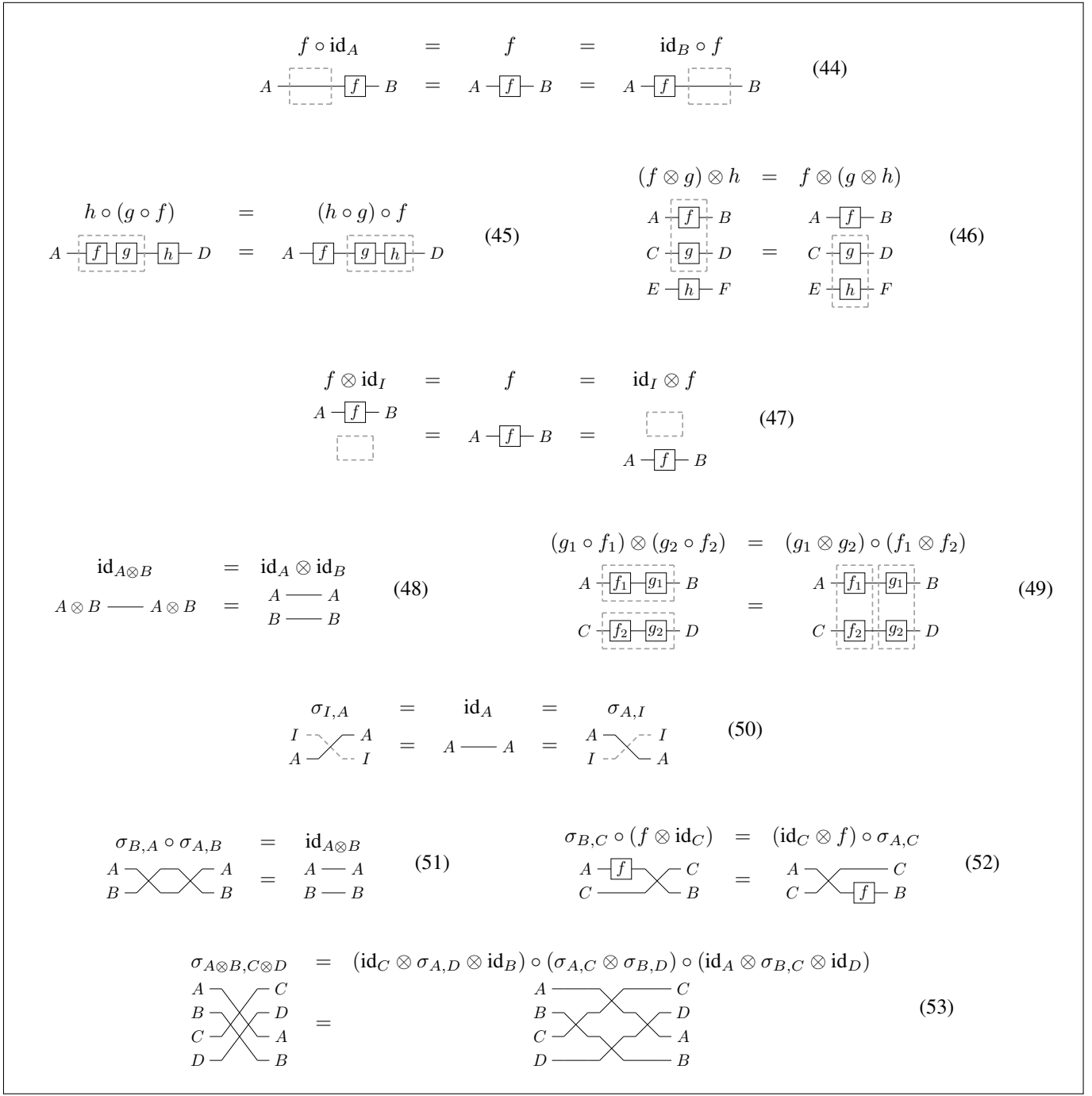
Diagrammatic reasoning has been very successful in recent years across various domains. These include quantum computing \cite{CD08,jeandel2018complete,zw,backens2018zh,AC04,CHMPV}, linear algebra \cite{sobocinski2015graphical,bonchi2014interacting}, control theory \cite{bonchi2014categorical,baez2014categories},  natural language processing \cite{coecke2021mathematics,zeng2016quantum} and symbolic dynamics \cite{jeandel2021strong} to cite a few. The natural mathematical framework for diagrammatic reasoning is that of symmetric monoidal categories, and more precisely, PROPs. This formalism allows representing processes and how they interact in space-time via sequential and parallel compositions.

Control is a central concept in quantum and reversible computational models. Roughly speaking, a transformation can be conditioned on an additional system such that, depending on its state, the transformation is either applied or not. For instance, controlling a NOT gate yields the well-known CNOT gate, and adding another control leads to the Toffoli gate. Notice that various quantum programming languages \cite{green2013quipper,fu2024proto,microsoft_qsharp,Qiskit,cirq} provide primitives that allow controlling gates or even circuits.

In this paper, we introduce a formalization of control as a constructor, i.e.~as a \emph{control functor} in a PROP: every diagram  $f$ with $n$ inputs and $n$ outputs yields a new diagram $\ctrl(f)$ with $1+n$ inputs and $1+n$ outputs, which represents its controlled version. A \emph{controlled PROP} is then  defined as a PROP equipped with a control functor. Similarly to PROPs, controlled PROPs can be used to define graphical languages by generators and relations.

The controlled PROP formalism is a versatile language that captures the fundamental properties of control and facilitates reasoning about processes. For instance, if an equation $f=g$ is derivable,  so is its controlled version $\ctrl(f)=\ctrl(g)$, whereas the latter can be cumbersome to derive when using the standard PROP formalism. Controlled PROPs rely on three constructors -- sequential composition, parallel composition, and control -- and specify how they interact. In particular, control does not preserve parallel composition, i.e., $\ctrl(f\otimes g)\neq \ctrl(f)\otimes \ctrl (g)$ since the latter acts on one more wire. However, it distributes naturally, as captured by the axioms of control functors.

We point out a fundamental property -- the \emph{conjugation law} -- which is central when dealing with reversing and control. Given $f$ and an invertible $g$, one can conjugate $f$ by $g$ with the \emph{compute-uncompute} pattern ${\invers {g}\circ f\circ g}$. This pattern is widely used in algorithm design, e.g.~for a basis change, for ancilla preparation and release, for oracle implementation, and also for circuit optimization \cite{fu2024proto}. Controlling such a pattern can be reduced to controlling only $f$, leading to the notion of \emph{conjugated PROP}, namely a controlled PROP satisfying the following so-called conjugation law.
\begin{equation*}
  \tf{conjugation_left}=\tf{conjugation_right}
\end{equation*}

An important motivation for our framework comes from quantum computing, where control is extensively used. Complete sets of relations for quantum circuits have been established in earlier works using the formalism of PROPs \cite{CHMPV,extensions,minimalQC}. These complete equational theories do not rely on control as a constructor, but point out the pivotal role of controlled operations in diagrammatic reasoning, and motivate the introduction of  controlled PROPs.  As our main application, we define the \emph{controllable quantum circuits} and give a complete set of relations for it. In its simplest form (see \cref{fig:axiomsintro}), where controllable quantum circuits are seen as a conjugated PROP, this complete set of relations is made of four basic equations (\cref{eq:2piintro,eq:additionintro,eq:swapintro,eq:hhintro}), together with the standard Euler decomposition (\cref{eq:eulerintro}). Notice that this equational theory is provably minimal. We also provide a complete set of equations for controllable circuits seen as a controlled PROP, with only a few additional relations corresponding to particular instances of the conjugation law.

\begin{figure*}[t]
  \fbox{\begin{minipage}{0.985\textwidth}\begin{center}
    \vspace{-1em}
    \hspace{-1.5em}
    \begin{subfigure}{0.20\textwidth}
      \begin{equation}\label{eq:2piintro}\tf{2pi}=\tf{empty}\end{equation}
    \end{subfigure}\hspace{2em}
    \begin{subfigure}{0.30\textwidth}
      \begin{equation}\label{eq:additionintro}\tf{alpha1alpha2}=\tf{alpha1+alpha2}\end{equation}
    \end{subfigure}\hspace{2em}
    \begin{subfigure}{0.30\textwidth}
      \begin{equation}\label{eq:hhintro}\tf{hh}=\tf{id}\end{equation}
    \end{subfigure}

    \begin{subfigure}{0.32\textwidth}
      \begin{equation}\label{eq:swapintro}\tf{swap}=\tf{swap-def_bare}\end{equation}
    \end{subfigure}
   
    \vspace{-0.3em}
    \hspace{-1.5em}
    \begin{subfigure}{0.81\textwidth}
      \begin{equation}\label{eq:eulerintro}\tf{euler-left_bare}=\tf{euler-right_bare}\end{equation}
    \end{subfigure}
    \vspace{0.4em}
  \end{center}\end{minipage}}
  \caption{\label{fig:axiomsintro} Complete set of relations for quantum circuits defined as a conjugated PROP generated by the Hadamard gate $\protect\tf{h}$ together with the global phases gates $\protect\tf{alpha}$ for any $\alpha\in\R$, where $\protect\tf{zalpha}$ is a notation for $\ctrl(\!\raisebox{0.5pt}{\protect\tf{alpha}}\!)$ and  $\protect\tf{not}$ is a notation for $\protect\tf{notdefz}$. See \cref{cor:ccqc} in \cref{sec:cqc} for details.}
\end{figure*}

Remarkably, in the complete set of relations we introduce for  controllable quantum circuits, all relations involve circuits acting on at most three qubits, whereas it has been shown in \cite{minimalQC} that any complete set of relations for (unitary) quantum circuits, defined as a PROP, requires at least one relation involving circuits acting on $n$ qubits for any $n\in\N$. Hence, controlled PROPs circumvent the unboundedness issue of PROPs, making them particularly well suited to diagrammatic reasoning.

As another application, we consider controllable quantum circuits with auxiliary qubits, a.k.a. \emph{ancillae}, for which we also introduce a simple complete set of relations. It illustrates that controlled PROPs are not necessarily made up only of endomorphisms (even if the control functor only applies to endomorphisms). Throughout the paper, we also give several examples illustrating the versatility of the formalism, including classical reversible circuits and qudit systems.

\subsection{Related work}

In graphical languages, control is a central notion that has been used in various contexts, but not as a constructor. For quantum circuits, the completeness procedure is based on a back-and-forth translation with optical circuits \cite{CHMPV} that extensively relies on controlled operations, and their basic algebra. In this context, controlled operations are considered as shorthand notations, as they can be decomposed into generators. In the ZX-calculus, controlled diagrams have also been extensively used, in particular for defining normal forms \cite{genericnf,AYY25,mcdowallrose2025}. Notice that the expressivity of the language yields a slightly different definition of control, allowing the control of arbitrary diagrams (not necessarily with the same number of inputs and outputs). Again in this context, controlled operations are considered as shorthand notations. Another interesting approach is to extend the quantum circuit model with a new kind of composition that graphically consists of vertically connecting  two gates, where the behavior of the resulting circuit depends on the Hamiltonians of the original ones. This work in progress allows, for instance, to represent a CNOT by connecting a Z gate and an X gate \cite{Schober24}. 

Additionally, various quantum programming languages (especially circuit description languages) allow for control of quantum circuits. In particular, Quipper supports control and a \emph{withcompute} primitive, which consists in  conjugating a circuit with another. These features  have been semantically formalized using controllable categories \cite{fu2024proto} which involve control functions -- not necessarily functors -- and \emph{withcompute} as a constructor,  to provide a categorical model for the high-level quantum programming language Quipper. Another quantum programming language introduced  recently in \cite{heunen25} allows any arbitrary unitaries to be generated from some simple global phases combined with a quantum analog of the \emph{if let} primitive, which can be embedded in our formalism (see \cref{sec:polycontrol}). $\sqrt\Pi$ \cite{squarerootspi} and Hadamard-$\Pi$ \cite{hadamardpi}  are two powerful quantum programming languages for approximately universal fragments of quantum computing, that are equipped with sound and complete equational theories. These languages differ from our approach in several ways: first, they consider fragments of unitary transformations, second, they are not graphical, and third, they rely on the direct sum as a primitive rather than control functors. Indeed, this promising approach based on rig categories relies on two constructors $\otimes$ and $\oplus$ corresponding to the tensor product and the direct sum respectively. These two products provide a means of expressing control and enable powerful decompositions of unitary transformations, which, however, cannot always be interpreted as graphical circuit-like transformations.  In contrast, control functors categorify the intuitive representation of control in circuits, in particular in the reversible and quantum cases. Finally, very recently, \cite{lemonnier2025} introduces a construction to turn PROPs with an identified involution into (poly)controlled PROPs with two control functors. These control functors are commutative and exhaustive\footnote{$\ctrl_2(g)\circ \ctrl_1(f)=\ctrl_1(f)\circ \ctrl_2(g)$ and $\ctrl_2(f)\circ \ctrl_1(f)=\textup{id}_1\otimes f$ (see \cref{def:commutativity,def:exhaustivity}).}. This recent pre-print relies on our completeness result on controlled quantum circuits (\cref{th:cqccompleteness}) to provide a variant of this result in their framework, among other completeness results for various fragments of quantum circuits.  Notice that the rig-based approach is currently specific to the binary case, whereas controlled PROPs are well-suited for more general settings (see Examples \ref{ex:hilbcontrolledprop} and \ref{ex:exhaust}).

\subsection{Outline of the paper}

In \cref{sec:background}, we fix the notations and recall the usual notions used to define graphical languages in the PROP formalism. In \cref{sec:control}, we introduce the controlled PROP formalism and its underlying notion of conjugation. In \cref{sec:qc}, we review how vanilla quantum circuits can be defined as a PROP and recall their already-known completeness result. In \cref{sec:cqc}, we define controllable quantum circuits as a controlled PROP, prove that they satisfy the conjugation law, and give a complete set of relations for them. In \cref{sec:aqc}, we add ancillae to the controllable quantum circuits and give a complete set of relations in these settings. In \cref{sec:polycontrol}, we define some notable properties that multiple coexisting control functors may satisfy, and give examples. In \cref{sec:conclusion}, we provide our concluding remarks. Additionally, the paper contains some appendices with detailed proofs.


\section{Background on PROPs and graphical languages}\label{sec:background}

A \emph{PROP} is, in category-theoretic terms, a strict symmetric monoidal category whose objects are generated by a single object, or equivalently, with $(\mathbb N, +)$ as a monoid of objects. Concretely, a PROP $\cat P$ consists of collections of morphisms $\cat P(n,m)$ for any objects $n,m\in \mathbb N$. We write $f:n\to m$ when $f$ is a morphism in $\cat{P}(n,m)$. Morphisms can be composed sequentially ($g\circ f:n\to m$, for any $f:n\to k$, $g:k\to m$) and in parallel ($f\otimes g:n+k\to m+\ell$, for any $f:n\to m$, $g:k\to \ell$). A PROP has specific morphisms: identities $\textup{id}_0:0\to 0$, $\textup{id}_1:1\to 1$ and a symmetry $\sigma_{1,1}: 2\to 2$. Identity on any object $n\in\N$ is inductively defined as $\textup{id}_{n}:= \textup{id}_{n-1}\otimes \textup{id}_1$ when $n>1$. Symmetry on any objects $n,m\in\N$ is inductively defined as $\sigma_{n,m}\defeq(\textup{id}_{1}\otimes\sigma_{n-1,m-1}\otimes\textup{id}_1)\circ(\sigma_{n-1,1}\otimes\sigma_{1,m-1})\circ(\textup{id}_{n-1}\otimes\sigma_{1,1}\otimes\textup{id}_{m-1})$ when $n\ne0\ne m$ and $\sigma_{n,0}\defeq\sigma_{0,n}\defeq\textup{id}_n$ otherwise.

PROPs have a nice graphical representation as string diagrams where objects are depicted as wires and morphisms as boxes. Here are some examples.
\begin{gather*}
    \begin{array}{cccccc}
        \textup{id}_0 \hspace{1em}&\hspace{1em} \textup{id}_1 \hspace{1em}&\hspace{1em} \sigma_{1,1} \hspace{1em}&\hspace{1em} f \hspace{1em}&\hspace{1em} g\circ f \hspace{1em}&\hspace{1em} f\otimes g\\
        \tf{empty} \hspace{1em}&\hspace{1em} \tf{id1} \hspace{1em}&\hspace{1em} \tf{swap} \hspace{1em}&\hspace{1em} \tf{f} \hspace{1em}&\hspace{1em} \tf{fg} \hspace{1em}&\hspace{1em} \tf{fgtensor}
    \end{array}
\end{gather*}

String diagrams take advantage of a second dimension by drawing the sequential composition horizontally and the parallel compositions vertically. Finally, the morphisms have to satisfy the \emph{coherence laws} of \cref{fig:coherenceprop}. Depicted graphically, many coherence laws are trivialized or correspond to diagram deformations. We denote by $\cat{P}_\textup{endo}$ the sub-PROP of endomorphisms of $\cat{P}$, defined as the collections of morphisms $\cat P(n,n)$ for any object $n\in \mathbb N$. 

\begin{example}\label{ex:hilbprop}
    Given $d\in\N$, let $\cat{FdHilb}_d$ be the PROP of qudit linear maps, meaning that ${\cat{FdHilb}_d(n,m) = \mathcal L(\mathbb C^{d^n},\mathbb C^{d^m})}$ is the collection of linear maps from $n$ qudits to $m$ qudits, and where the parallel composition is the usual tensor product. Moreover, let $\cat{Iso}$ be the sub-PROP of $\cat{FdHilb}_2$ restricted to isometries\footnote{$f:n\to m$ is an isometry if its adjoint $f^\dagger:m\to n$ is its left inverse, i.e. $f^\dagger \circ f=\textup{id}_n$.}, and $\cat{Qubit}:= \cat{Iso}_\textup{endo}$ its restriction to unitary evolutions. 
\end{example}

Given two PROPs $\cat{P}$ and $\cat{Q}$, a \emph{functor} ${F:\cat{P}\to\cat{Q}}$ is a map that assigns to each object $n$ of $\cat{P}$ an object $F(n)$ of $\cat{Q}$, and to each morphism ${f\in \cat{P}(n,m)}$ a morphism ${F(f)\in\cat{Q}(F(n),F(m))}$ while preserving the compositional structure, meaning ${F(\textup{id}_0)=\textup{id}_{F(0)}}$, ${F(\textup{id}_1)=\textup{id}_{F(1)}}$, and ${F(g\circ f)=F(g)\circ F(f)}$ whenever ${g\circ f}$ is defined. We say that $F$ is \emph{monoidal} when it also preserves the parallel composition, meaning ${F(0)=0}$, ${F(n+m)=F(n)+F(m)}$ and $F(f\otimes g)=F(f)\otimes F(g)$. A \emph{PROP functor} is a monoidal functor that also preserves the symmetry, meaning ${F(\sigma_{1,1})=\sigma_{F(1),F(1)}}$. A \emph{dagger functor} ${(\cdot)^\dagger:\cat{P}\to\cat{P}}$ is an identity-on-object involutive contravariant PROP functor, meaning that every morphism ${f:n\to m}$ has a dagger ${f^\dagger:m\to n}$ satisfying $f^{\dagger\dagger}=f$, $(g\circ f)^\dagger=f^\dagger\circ g^\dagger$, $(f\otimes g)^\dagger=f^\dagger\otimes g^\dagger$ and additionally $\sigma_{1,1}^\dagger=\sigma_{1,1}$.

Graphical languages, such as Boolean or quantum circuits, can be defined as PROPs by generators and relations. That is, the diagrams (or morphisms) are generated inductively by sequential and parallel compositions of generators, quotiented by the smallest congruence\footnote{A congruence is an equivalence relation $\mathcal R$ on the set of morphisms such that if $f_1 \mathcal R f_2$ and $g_1 \mathcal R g_2$, then $(g_1\circ f_1) \mathcal R (g_2\circ f_2)$ and $(f_1\otimes g_1) \mathcal R (f_2\otimes g_2)$.} that satisfies the set of relations as well as the coherence laws of PROPs. Given a graphical language $\cat{P}$, the corresponding congruence is denoted  ${\cat{P}\vdash D_1=D_2}$ where $D_1$ and $D_2$ are two diagrams that can be transformed one into the other using the relations and the coherence laws.

A graphical language $\cat{P}$ often comes with an interpretation (or semantics) functor ${\interp{\cdot}:\cat{P}\to\cat{S}}$. The graphical language $\cat{P}$ is universal when $\interp{\cdot}$ is full, i.e. for any morphism $f$ in $\cat{S}$, there exists a diagram $D$ in $\cat P$ such that ${\interp D =f}$; and $\cat P$ is complete when $\interp{\cdot}$ is faithful, i.e. ${\interp{D_1}=\interp{D_2}}$ implies ${\cat{P}\vdash D_1=D_2}$ for any ${D_1,D_2\in\cat{P}(n,m)}$. We say that $\cat{P}$ is universally complete when $\interp{\cdot}$ is full and faithful. Completeness guarantees that all fundamental properties of the semantic domain  $\cat{S}$ are graphically captured by $\cat{P}$.

\begin{figure*}[t]
    \fbox{\begin{minipage}{0.985\textwidth}\begin{center}
        \vspace{-0.5em}
        \hspace{-1.5em}
        \begin{subfigure}{0.59\textwidth}
            \begin{equation}\label{eq:seqidentity}
                \begin{array}{ccccc}
                    f\circ \textup{id}_n&=&f&=&\textup{id}_m\circ f\\[.4em]
                    \tf{seqidentity_left}&=&\tf{seqidentity_mid}&=&\tf{seqidentity_right}
                \end{array}
            \end{equation}
        \end{subfigure}

        \vspace{0em}
        \hspace{-1.5em}
        \begin{subfigure}{0.52\textwidth}
            \begin{equation}
                \begin{array}{ccc}
                    h\circ(g\circ f)&=&(h\circ g)\circ f\\[.4em]
                    \tf{seqassociativity_left}&=&\tf{seqassociativity_right}
                \end{array}
            \end{equation}
        \end{subfigure}\hspace{2em}
        \begin{subfigure}{0.41\textwidth}
            \begin{equation}
                \begin{array}{ccc}
                    (f\otimes g)\otimes h&=&f\otimes(g\otimes h)\\[.4em]
                    \tf{parassociativity_left}&=&\tf{parassociativity_right}
                \end{array}
            \end{equation}
        \end{subfigure}

        \vspace{0em}
        \hspace{-1.5em}
        \begin{subfigure}{0.47\textwidth}
            \begin{equation}\label{eq:paridentity}
                \begin{array}{ccccc}
                    f\otimes \textup{id}_0&=&f&=&\textup{id}_0\otimes f\\[.4em]
                    \tf{paridentity_left}&=&\tf{paridentity_mid}&=&\tf{paridentity_right}
                \end{array}
            \end{equation}
        \end{subfigure}

        \vspace{0em}
        \hspace{-1.5em}\begin{subfigure}{0.60\textwidth}
            \begin{equation}\label{eq:interchange}
                \begin{array}{ccc}
                    (g_1\circ f_1)\otimes(g_2\circ f_2)&=&(g_1\otimes g_2)\circ(f_1\otimes f_2)\\[.4em]
                    \tf{interchange_left}&=&\tf{interchange_right}
                \end{array}
            \end{equation}
        \end{subfigure}

        \vspace{0em}
        \hspace{-1.5em}
        \begin{subfigure}{0.38\textwidth}
            \begin{equation}\label{eq:involution}
                \begin{array}{ccc}
                    \sigma_{1,1}\circ\sigma_{1,1}&=&\textup{id}_{2}\\[.4em]
                    \tf{involution_left}&=&\tf{involution_right}
                \end{array}
            \end{equation}
        \end{subfigure}\hspace{2em}
        \begin{subfigure}{0.52\textwidth}
            \begin{equation}\label{eq:naturality}
                \begin{array}{ccc}
                    \sigma_{m,1}\circ(f\otimes\textup{id}_1)&=&(\textup{id}_1\otimes f)\circ\sigma_{n,1}\\[.4em]
                    \tf{naturality_left}&=&\tf{naturality_right}
                \end{array}
            \end{equation}
        \end{subfigure}
        \vspace{-0.2em}
    \end{center}\end{minipage}}
    \caption{\label{fig:coherenceprop}Coherence laws of PROPs. Notice that the dotted boxes are just here to highlight the correspondence with the syntax.}
\end{figure*}


\section{Control and conjugation}\label{sec:control}

In this section, we formalize the notion of control as a constructor in a PROP. Formally, this is achieved by equipping a PROP with a control functor that satisfies some coherence laws.

\begin{definition}[control functor]\label{def:controlfunctor}
    Given a PROP $\cat{P}$, a \emph{control functor} is a functor ${\ctrl:\cat{P}_\textup{endo}\to\cat{P}_\textup{endo}}$ mapping any object ${n\in\N}$ to the object ${1+n}$ and mapping any morphism ${f:n\to n}$ to a morphism ${\ctrl (f):(1+n)\to(1+n)}$ while satisfying the coherence laws depicted in \cref{fig:coherencecprop}.
\end{definition}

\begin{figure*}[t]
    \fbox{\begin{minipage}{0.985\textwidth}\begin{center}
        \vspace{-0.5em}
        \hspace{-0.4em}
        \begin{subfigure}{0.36\textwidth}
            \begin{equation}\label{eq:strenght}
                \begin{array}{ccc}
                    \scalebox{0.90}{$\ctrl(f\otimes \textup{id}_1)$}&=&\scalebox{0.90}{$\ctrl(f)\otimes \textup{id}_1$}\\[.4em]
                    \tf{strenght_left}&=&\tf{strenght_right}
                \end{array}
            \end{equation}
        \end{subfigure}\hspace{1.8em}
        \begin{subfigure}{0.58\textwidth}
            \begin{equation}\label{eq:controlswap}
                \begin{array}{ccc}
                    \scalebox{0.90}{$\ctrl(\ctrl(f))\circ(\sigma_{1,1}\otimes \textup{id}_{n})$}&=&\scalebox{0.90}{$(\sigma_{1,1}\otimes \textup{id}_{n})\circ\ctrl(\ctrl (f))$}\\[.4em]
                    \tf{ctrlswap_left}&=&\tf{ctrlswap_right}
                \end{array}
            \end{equation}
        \end{subfigure}\hspace{3em}

        \vspace{-1em}
        \hspace{-0.5em}
        \begin{equation*}
            \scalebox{0.90}{$\ctrl((\textup{id}_{k}\otimes \sigma_{1,1}\otimes \textup{id}_{\ell})\circ f\circ(\textup{id}_{k}\otimes \sigma_{1,1}\otimes \textup{id}_{\ell}))=(\textup{id}_{k+1}\otimes \sigma_{1,1}\otimes \textup{id}_{\ell})\circ \ctrl(f)\circ(\textup{id}_{k+1}\otimes \sigma_{1,1}\otimes \textup{id}_{\ell})$}
        \end{equation*}\vspace{-1.5em}
        \begin{equation}\label{eq:swapconjugation}
            \tf{swapconjugation_right}\hspace{4em}=\hspace{4em}\tf{swapconjugation_left}\hspace{2.5em}
        \end{equation}
    \vspace{0em}
    \end{center}\end{minipage}}
    \caption{\label{fig:coherencecprop}Coherence laws of control functors. \cref{eq:strenght,eq:controlswap} are defined for any $n\in\N$ and any $f\in\cat{P}(n,n)$. \cref{eq:swapconjugation} is defined for any $k,\ell\in\N$ and any $f\in\cat{P}(k+2+\ell,k+2+\ell)$. Several wires are sometimes depicted as a single wire for simplicity.}
\end{figure*}

Graphically, functors can be depicted using boxes \cite{mellies2006functorial}. Thus, $\ctrl (f)$ is depicted as a bullet point on the additional wire connected to the diagram $f$, enclosed within a dotted box. This dotted box is omitted  when the diagram is made of a single morphism or when there are nested controls (thanks to \cref{eq:controlswap}).
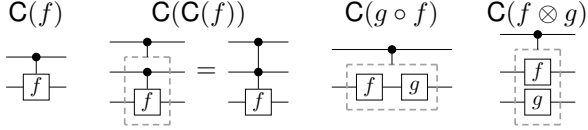
\begin{gather*}
    \begin{array}{cccc}
        \ctrl (f) \hspace{2em}&\hspace{2em} \ctrl(\ctrl (f)) \hspace{2em}&\hspace{2em} \ctrl (g\circ f) \hspace{2em}&\hspace{2em} \ctrl (f\otimes g) \\
        \tf{ctrlf} \hspace{2em}&\hspace{2em} \tf{ctrlctrlfvar}\!=\!\tf{ctrlctrlf} \hspace{2em}&\hspace{2em} \tf{ctrlfseqg} \hspace{2em}&\hspace{2em} \tf{ctrlfparg}
    \end{array}
\end{gather*}

We may write $\ctrl^n(f)$ for $n$ nested applications of the control functor $\ctrl$ from $f$. Formally, $\ctrl^0(f)\defeq f$ and $\ctrl^{n+1}(f)\defeq\ctrl(\ctrl^{n}(f))$. Also, the coherence laws of PROPs (\cref{fig:coherenceprop}) allow us to simplify the drawings by stretching the line to depict morphisms that are controlled by several non-consecutive wires. The following diagrams are the same modulo the coherence laws.

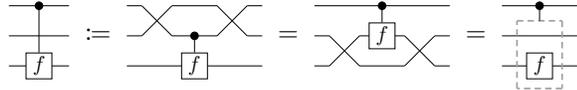
\begin{equation*}
    \tf{swapctrlf_01}
    \defeq\tf{swapctrlf_02}
    =\tf{swapctrlf_03}
    =\tf{swapctrlf_05}
\end{equation*}

The coherence laws of control functors capture the basic behavior of a control: identities are not affected by control (\cref{eq:strenght}), nested controls commute (\cref{eq:controlswap}), and finally, the  symmetry, which is left-invertible by definition in a PROP, satisfies the conjugation law (\cref{eq:swapconjugation}). Notice that the functoriality of $\ctrl$ additionally implies $\ctrl(g\circ f)=\ctrl(g)\circ\ctrl(f)$ (whenever $g\circ f$ is defined), and $\ctrl(\textup{id}_n)=\textup{id}_{1+n}$ (in particular, $\ctrl(\raisebox{1pt}{\tinytf{empty}})=\raisebox{1pt}{\tinytf{id}}$). However, $\ctrl$ is not monoidal because $\ctrl(f\otimes g)\ne\ctrl(f)\otimes\ctrl(g)$ in general, but the expected behavior of the control functor on parallel compositions is provided by the coherence laws of control functors, as illustrated in the following derivation.

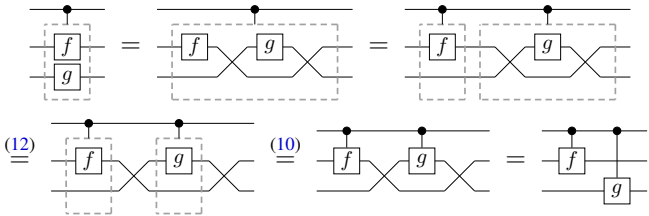
\begin{multline*}
    \tf{ctrlpar_01}
    =\tf{ctrlpar_02}
    =\tf{ctrlpar_03}\\[.4em]
    \eqoneref{eq:swapconjugation}\tf{ctrlpar_04}
    \eqoneref{eq:strenght}\tf{ctrlpar_05}
    =\tf{ctrlpar_06}
\end{multline*}

\begin{example}\label{ex:hilbcontrolledprop}
    In $\cat{FdHilb}_2$, the standard notion of control is to apply a linear map ${U\in \mathcal L(\C^{2^n},\C^{2^n})}$ on the targeted system when the control qubit is in the state $\ket{1}$, and to apply the identity $I_{2^n}$ when the control qubit is in the state $\ket{0}$. This can be embedded as the following control functor.
    \begin{equation*}
        \ctrl_{\ket{1}}:U\mapsto\ketbra{0}{0}\otimes I_{2^n}+\ketbra{1}{1}\otimes U
    \end{equation*}
    However, this is not the only control functor. For instance, 
    \begin{equation*}
        \ctrl_{\ket{0}}:U\mapsto\ketbra{0}{0}\otimes U+\ketbra{1}{1}\otimes I_{2^n}
        \hspace{1em}\text{and}\hspace{1em}
        \ctrl_{\ket{-}}:U\mapsto\ketbra{+}{+}\otimes I_{2^n}+\ketbra{-}{-}\otimes U
    \end{equation*}
    are also control functors in $\cat{FdHilb}_2$ (where $\ket{+}\defeq\nicefrac{1}{\sqrt{2}}(\ket{0}+\ket{1})$ and $\ket{-}\defeq\nicefrac{1}{\sqrt{2}}(\ket{0}-\ket{1})$). In higher dimensions, $\cat{FdHilb}_d$ admits a control functor 
    \begin{equation*}
        \textstyle\ctrl_{\ket{k}}: U\mapsto\ketbra{k}{k}\otimes U+\sum_{\ell\ne k}\ketbra{\ell}{\ell}\otimes I_{d^n}
    \end{equation*}
    for any ${0\le k<d}$. Notice however that the  map ${U\mapsto\sum_{k}\ketbra{k}{k}\otimes U^k}$ is not a control functor as it fails to be functorial in general when $d>2$.
\end{example}

We now focus on PROPs equipped with a {distinguished} control functor, that we call \emph{controlled PROPs}. The general case of PROPs with several control functors is described in \cref{sec:polycontrol}. Notice that a single control functor already offers a rich structure as other control functors can be obtained by {conjugating} the control wire with some invertible morphism. For instance in $\cat{FdHilb}_2$ one can recover the various examples described in \cref{ex:hilbcontrolledprop} by conjugating the control qubit with NOT or Hadamard gates.

\begin{definition}[controlled PROP]\label{def:controlledprop}
    A \emph{controlled PROP} is a PROP equipped with a  control functor.
\end{definition}

Throughout the paper, we denote by $\ctrl$ the control functor of a controlled PROP, unless a different notation is specified.

Similarly to PROPs, a controlled PROP $\cat{P}$ can be defined by generators and relations. In this setting, the diagrams are inductively defined as follows: the generators are morphisms, and for any morphisms ${f_1:n\to k}$, ${g_1:k\to m}$, ${f_2:n\to m}$, ${g_2:k\to\ell}$ and ${f:n\to n}$, we have that ${g_1\circ f_1}$, ${f_2\otimes g_2}$ and ${\ctrl(f)}$ are morphisms. The morphisms are then quotiented by the smallest congruence ${\cat P \vdash \cdot = \cdot}$ that satisfies the given relations as well as the coherence laws in \cref{fig:coherenceprop} and \cref{fig:coherencecprop}, where the congruence does not only preserve the parallel and sequential compositions but also the control, i.e. if ${\cat{P}\vdash f=g}$ then ${\cat P \vdash \ctrl(f)=\ctrl(g)}$. In other words, if an equation $f=g$ can be derived, its controlled version ${\ctrl(f)=\ctrl(g)}$ follows directly by construction.
\begin{example}
    We consider the controlled PROP $\cat{CNOT}$ generated by $\raisebox{1pt}{\smalltf{not-s}}:1\to 1$ with relations $\raisebox{1pt}{\smalltf{notnotsmall}}=\raisebox{1pt}{\smalltf{id}}$, $\raisebox{2pt}{\smalltf{cnotswapL}}=\raisebox{2pt}{\smalltf{cnotswapR}}$ and $\raisebox{2pt}{\smalltf{notcnotnot}}=\raisebox{2pt}{\smalltf{cnotnot}}$. Notice that $\raisebox{2pt}{\smalltf{cnotsmall}}$ is not a generator but a diagram obtained by applying the control functor $\ctrl$ to $\raisebox{1pt}{\smalltf{not-s}}$, moreover the equation $\raisebox{2pt}{\smalltf{cnotcnotsmall}}=\raisebox{2pt}{\smalltf{id2small}}$ can be derived. 
\end{example}

\begin{definition}[conjugated PROP]
    A \emph{conjugated PROP} $\cat{P}$ is a controlled PROP such that \cref{eq:conjugation} is satisfied for any $f\in\cat{P}(m,m)$, $g\in\cat{P}(n,m)$, and $h\in\cat{P}(m,n)$ such that ${\cat{P}\vdash h\circ g=\textup{id}_n}$.
    \begin{equation}\label{eq:conjugation}
        \begin{array}{ccc}
            \ctrl(h\circ f\circ g)&=&(\textup{id}_1\otimes h)\circ \ctrl(f)\circ (\textup{id}_1\otimes g)\\[.4em]
            \tf{conjugation_gfh_left}&=&\tf{conjugation_gfh_right}
        \end{array}
    \end{equation}
\end{definition}

\cref{eq:conjugation} is known as the \emph{conjugation law}. This law is particularly important in algorithm design where the \emph{compute-uncompute} pattern $\invers g\circ f\circ g$ is often used to prepare and release ancillary bits (or qubits) and implement oracles. This law can be used as much as possible in some compilation processes to minimize the scope of control as explained in \cite{fu2024proto}.

\begin{example}
    $\cat{FdHilb}_2$ with $\ctrl_{\ket{1}}$ as control functor is a conjugated PROP.
\end{example}

Similarly to controlled PROPs, conjugated PROPs can be defined with generators and relations. The morphisms are freely generated from the generators using sequential compositions, parallel compositions and control, and then quotiented by the given relations together with the coherence laws of \cref{fig:coherenceprop} and \cref{fig:coherencecprop} as well as the conjugation law (\cref{eq:conjugation}). Notice that \cref{eq:conjugation} is somehow circular as it applies to morphisms with a left inverse; as a consequence, one can define a sequence of congruences $(\cat P\vdash_k \cdot=\cdot)_{k\in \mathbb N}$ as follows: ${\cat P\vdash_0 \cdot =\cdot}$ is the congruence of the corresponding controlled PROP (i.e. using all equations but the conjugation law), and ${\cat P\vdash_{k+1} \cdot =\cdot}$ is the minimal congruence of the controlled PROP augmented with the equations $\ctrl(h\circ f\circ g)=(\textup{id}_1\otimes h)\circ \ctrl (f)\circ (\textup{id}_1\otimes g)$ for any $f:m\to m$, $g:n\to m$, and $h :m\to n$ such that $\cat P\vdash_k h\circ g = \textup{id}_n$. Tarski's fixed point theorem guarantees the existence of a fixed point congruence $\cat P\vdash \cdot= \cdot$.

Notice that any PROP admits two trivial control functors ${f\mapsto \textup{id}_1\otimes f}$ and ${f\mapsto \textup{id}_{1+n}}$ (when ${f:n\to n}$) which correspond to the two degenerate cases of a control, namely when $f$ is respectively always or never applied.  The notion of \emph{points}, which we introduce below, can be used to rule out these two degenerate  cases. Intuitively, the points of a control functor are two specific morphisms \emph{{false}} (${\tf{winit}:0\to 1}$) and \emph{true} (${\tf{binit}:0\to 1}$) that respectively fire and annihilate the controlled operation when they are plugged to the control wire.

\begin{definition}[points]
    A control functor ${\ctrl:\cat{P}_\textup{endo}\to\cat{P}_\textup{endo}}$ has \emph{points} if there exist two morphisms ${\tf{winit}:0\to 1}$ and ${\tf{binit}:0\to 1}$ such that the following equations are satisfied for any ${f\in\cat{P}(n,n)}$.
    \begin{gather*}
        \begin{array}{ccc}
            \ctrl(f)\circ(\tf{winit}\otimes \textup{id}_n) &=& \tf{winit}\otimes \textup{id}_n\\[.4em]
            \tf{winitcf} &=& \tf{winitidf}
        \end{array} \hspace{4em}
        \begin{array}{ccc}
            \ctrl(f)\circ(\tf{binit}\otimes \textup{id}_n) &=& \tf{binit}\otimes f\\[.4em]
            \tf{binitcf} &=& \tf{binitf}
        \end{array}
    \end{gather*}
\end{definition}

\begin{example}
    In $\cat{FdHilb}_2$, $\ket{0}$ and $\ket{1}$ are points of the control functor $\ctrl_{\ket{1}}$. In $\cat{FdHilb}_d$, $\ket{\ell}$ and $\ket{k}$ are points of the control functor $\ctrl_{\ket{k}}$ whenever $\ell\ne k$.
\end{example}

Notice however that points do not necessarily exist, in particular when all morphisms of the PROP are endomorphisms, like in $\cat{Qubit}$. In \cref{sec:cqc}, we consider the controlled PROP of quantum circuits without ancillae, where all generators are endomorphisms, and which has thus no points, whereas \cref{sec:aqc} is dedicated to quantum circuits with ancillae, leading to controlled PROPs with points. We first review the properties of quantum circuits defined as a PROP in the following section.


\section{Vanilla quantum circuits}\label{sec:qc}

In this section, following  \cite{CHMPV,minimalQC}, we consider vanilla quantum circuits defined as a PROP (without control functor) by generators and relations and interpreted as unitary maps in $\cat{Qubit}$ (see \cref{ex:hilbprop}). We also recall the already-known completeness result.

\begin{definition}[vanilla quantum circuits]\label{def:qc}
  Let $\cat{QC}$ be the PROP of \emph{vanilla quantum circuits} generated by the following generators (where ${\alpha\in\R}$) and the relations $\et{R}_{\textup{v}}$ depicted in \cref{fig:qcaxioms}.
  \vspace{-0.5em}
  \begin{gather*}
    \!\tf{alpha}\!:0\to0 \hspace{3em}
    \tf{h}:1\to 1 \hspace{3em}
    \tf{zalpha}:1\to1 \hspace{3em}
    \tf{cnot}:2\to2
  \end{gather*}
\end{definition}

\begin{definition}[interpretation]\label{def:stdinterp}
  Let ${\interp{\cdot}_{\textup{v}}:\cat{QC}\to\cat{Qubit}}$ be the interpretation of vanilla quantum circuits inductively defined as the following identity-on-object PROP functor.
  \vspace{-0.5em}
  \begin{gather*}
    \interp{\tf{h}}_{\textup{v}}\defeq \ketbra{+}{0}+\ketbra{-}{1} \hspace{4em}
    \interp{\tf{zalpha}}_{\textup{v}}\defeq \ketbra{0}{0}+e^{i\alpha}\ketbra{1}{1} \\[0.5em]
    \interp{\tf{alpha}}_{\textup{v}}\defeq e^{i\alpha} \hspace{4em}
    \textstyle\biginterp{\tf{cnot}}_{\textup{v}}\defeq \sum_{x,y\in\{0,1\}}\ketbra{x,x\oplus y}{x,y}
  \end{gather*}
\end{definition}

The functor ${\interp{\cdot}_{\textup{v}}:\cat{QC}\to\cat{Qubit}}$ assigns a unitary to all vanilla quantum circuits. Conversely, it is known that any unitary can be expressed as a vanilla quantum circuit. Thus, $\cat{QC}$ is universal for $\cat{Qubit}$, meaning that $\interp{\cdot}_{\textup{v}}$ is full.

\begin{figure*}[t]
  \fbox{\begin{minipage}{0.985\textwidth}\begin{center}
    \vspace{-1.2em}
    \hspace{-1em}
    \begin{subfigure}{0.22\textwidth}
      \begin{equation}\label{eq:2pibare}\tf{2pi}=\tf{empty}\end{equation}
    \end{subfigure}\hspace{1.5em}
    \begin{subfigure}{0.33\textwidth}
      \begin{equation}\label{eq:additionbare}\tf{alpha1alpha2}=\tf{alpha1+alpha2}\end{equation}
    \end{subfigure}\hspace{1.5em}
    \begin{subfigure}{0.32\textwidth}
    \begin{equation}\label{eq:hhbare}\tf{hh}=\tf{id}\end{equation}
    \end{subfigure}
   
    \hspace{-1em}
    \begin{subfigure}{0.29\textwidth}
      \begin{equation}\label{eq:0bare}\tf{c0_bare}=\tf{id}\end{equation}
    \end{subfigure}\hspace{3em}
    \begin{subfigure}{0.48\textwidth}
      \begin{equation}\label{eq:cadditionbare}\tf{calpha1calpha2_bare}=\tf{calpha1+alpha2_bare}\end{equation}
    \end{subfigure}

    \hspace{-1em}
    \begin{subfigure}{0.82\textwidth}
      \begin{equation}\label{eq:eulerbare}\tf{euler-left_bare}=\tf{euler-right_bare}\end{equation}
    \end{subfigure}

    \hspace{-1em}
    \begin{subfigure}{0.38\textwidth}
      \begin{equation}\label{eq:zcommutbare}\tf{cnotzalphacnot}=\tf{zalphaid}\end{equation}
    \end{subfigure}\hspace{3em}
    \begin{subfigure}{0.33\textwidth}
      \begin{equation}\label{eq:swapbare}\tf{swap}=\tf{swap-def_bare}\end{equation}
    \end{subfigure}

    \hspace{-1em}
    \begin{subfigure}{0.41\textwidth}
      \begin{equation}\label{eq:czbare}\tf{hcnoth}=\tf{ccpibox}\end{equation}
    \end{subfigure}\hspace{3em}
    \begin{subfigure}{0.35\textwidth}
      \begin{equation}\label{eq:mc2pibare}\tf{mc2pi}=\tf{mid}\end{equation}
    \end{subfigure}
    \vspace{0em}
  \end{center}\end{minipage}}
  \caption{\label{fig:qcaxioms} Relations $\et{R}_{\textup{v}}$ of $\cat{QC}$, containing an instance of \cref{eq:mc2pibare} for any number of qubits $n\ge3$, and where $\lambda^n(\alpha)$ is defined by \cref{eq:lambdadef}. The relations between $\alpha_1,\alpha_2$ and $\beta_0,\beta_1,\beta_2,\beta_3$ are explained in \cref{sec:qc}.}
\end{figure*}

In particular, we can implement any controlled operation by composing basic gates. For instance, the multi-controlled Z-rotations can be implemented for any ${\alpha\in\R}$ and ${n\in\N}$ by the circuit ${\lambda^n(\alpha)\in\cat{QC}(n,n)}$ inductively defined as follows.
\begin{equation}\label{eq:lambdadef}
  \begin{split}
    \lambda^0(\alpha)\defeq \tf{alpha} \hspace{3em}
    \lambda^1(\alpha)\defeq\tf{zalpha}  \hspace{6em}\\[0.5em]
    \smalltf{mcalpha}\defeq\smalltf{mcalpha-induc-def}
  \end{split}
\end{equation}

We can check that its interpretation (see \cref{def:stdinterp}) is the desired Z-rotation with parameter $\alpha$ and controlled by the first $n-1$ qubits.
\begin{equation*}
  \textstyle\interp{\lambda^n(\alpha)}_{\textup{v}}=e^{i\alpha}\ketbra{1\dots1}{1\dots1}+\sum_{x\ne1\dots1}\ketbra{x}{x}
\end{equation*}

As a consequence, each instance of \cref{eq:mc2pibare} is semantically trivial ($ {\textstyle\interp{\lambda^n(2\pi)}_{\textup{v}} = I_{2^n}}$) but syntactically involved as the circuit ${\lambda^n(\pi)}$ contains a number of \tf{zpiover2nminus1} gates which is exponential in $n$. 

It has been shown in \cite{minimalQC} that $\cat{QC}$ is complete.
\begin{theorem}[\cite{minimalQC}]\label{th:qccompleteness}
  The PROP $\cat{QC}$ is universally complete for $\cat{Qubit}$, meaning that the functor $\interp{\cdot}_{\textup{v}}:\cat{QC}\to\cat{Qubit}$ is full and faithful.
\end{theorem}

All the relations in $\et{R}_{\textup{v}}$, except \cref{eq:mc2pibare}, are fairly simple and commonly used in the literature. Surely the most powerful relation is \cref{eq:eulerbare}. This relation follows from the well-known Euler decomposition which states that any unitary can be decomposed, up to a global phase, into basic X- and Z-rotations. The angles $\alpha_1,\alpha_2$ are arbitrary in $\R$ whereas the angles $\beta_0,\beta_1,\beta_2,\beta_3$ are restricted to $[0,2\pi)$ and are computed by explicit functions as follows using the intermediate complex numbers $u,v\in\C$.
\begin{gather*}
    \begin{array}{l}
      u\defeq -\sin\left(\nicefrac{\alpha_1+\alpha_2}{2}\right)+i\cos\left(\nicefrac{\alpha_1-\alpha_2}{2}\right)\\
      v\defeq +\cos\left(\nicefrac{\alpha_1+\alpha_2}{2}\right)-i\sin\left(\nicefrac{\alpha_1-\alpha_2}{2}\right)
    \end{array}
    \;\;\;\text{and}\;\;\;
    \beta_0\defeq\frac{(\pi+\alpha_1+\alpha_2-\beta_1-\beta_2-\beta_3)}{2}
  \end{gather*}
\begin{center}
  \begin{tabular}{c|c|c|c}
    & $\beta_1$ & $\beta_2$ & $\beta_3$ \\
    \hline
    if $v=0$ & $2\arg(u)$ & $0$ & $0$\\
    \hline
    if $u=0$ & $2\arg(v)$ & $\pi$ & $0$\\
    \hline
    otherwise \hspace{0.2em}&\hspace{0.2em} $\arg(u)+\arg(v)$ \hspace{0.2em}&\hspace{0.2em} $2\arg\left(i+\left\lvert\nicefrac{u}{v}\right\rvert\right)$ \hspace{0.2em}&\hspace{0.2em} $\arg(u)-\arg(v)$
  \end{tabular}
\end{center}

It has been shown in \cite{minimalQC} that every relation of $\et{R}_{\textup{v}}$ is necessary. This means that none can be removed without losing completeness. Notice that there is an instance of \cref{eq:mc2pibare} for any number of qubits $n\ge3$. All such instances are necessary. More generally, any set of relations yielding a universally complete graphical language (defined as a PROP) for unitary maps requires at least one relation acting on $n$ qubits for all $n\in\N$. As we will see in the following, the use of controlled PROPs simplifies the diagrammatic reasoning on quantum circuits, and is, in particular, a way to go around the necessity of relations acting on an unbounded number of qubits.


\section{Controllable quantum circuits}\label{sec:cqc}

In this section we define the \emph{controllable quantum circuits} as a controlled PROP, prove that the conjugation law holds in these settings, and give a completeness result for it. Using the controlled PROP formalism, we can use much simpler gate sets to define quantum circuits. This is because a Z-rotation gate can be seen as a controlled global phase gate, and a CNOT gate can be seen as a controlled NOT gate. In fact, we can just take the global phases together with the Hadamard gate as a gate set and still capture all unitaries in $\cat{Qubit}$.

\begin{definition}[controllable quantum circuits]
  Let $\cat{CQC}$ be the controlled PROP of \emph{controllable quantum circuits} generated by the following generators (where $\alpha\in\R$) and the relations $\et{R}_{\textup{c}}$ depicted in \cref{fig:cqcaxioms}.
  \vspace{-0.5em}
  \begin{gather*}
    \!\tf{alpha}\!:0\to0 \hspace{3em}
    \tf{h}:1\to 1
  \end{gather*}
\end{definition}

There are two kinds of relations in $\et{R}_{\textup{c}}$: On the one hand, \cref{eq:2pi,eq:addition,eq:swap,eq:hh,eq:euler} directly correspond to some relations in $\et{R}_{\textup{v}}$. In particular the parameters of \cref{eq:euler} are computed with the same functions as in \cref{eq:eulerbare}. Notice that some relations of $\et{R}_{\textup{v}}$ do not appear in $\et{R}_{\textup{c}}$ as they are trivialised within the controlled PROP formalism. On the other hand, \cref{eq:conjhcalpha,eq:conjcalphah,eq:conjhccpi,eq:conjccpih} are new and are all instances of the conjugation law.

We extend the interpretation of vanilla quantum circuits (\cref{def:stdinterp}) to embed the control. This is achieved with the following PROP functor.
\begin{definition}[interpretation]
  Let ${\interp{\cdot}_{\textup{c}}:\cat{CQC}\to\cat{Qubit}}$ be the interpretation of controllable quantum circuits inductively defined as the following identity-on-object PROP functor.
  \vspace{-0.5em}
  \begin{gather*}
    \interp{\tf{alpha}}_{\textup{c}}\defeq e^{i\alpha} \hspace{4em}
    \interp{\tf{h}}_{\textup{c}}\defeq \ketbra{+}{0}+\ketbra{-}{1} \\[0.5em]
    \interp{\ctrl (C)}_{\textup{c}}\defeq \ketbra{0}{0}\otimes I_{2^n}+\ketbra{1}{1}\otimes \interp{C}_{\textup{c}}
  \end{gather*}
\end{definition}

\begin{figure*}[t]
  \fbox{\begin{minipage}{0.985\textwidth}\begin{center}
    \vspace{-0.5em}
    \hspace{-1.5em}
    \begin{subfigure}{0.22\textwidth}
      \begin{equation}\label{eq:2pi}\tf{2pi}=\tf{empty}\end{equation}
    \end{subfigure}\hspace{1.5em}
    \begin{subfigure}{0.33\textwidth}
      \begin{equation}\label{eq:addition}\tf{alpha1alpha2}=\tf{alpha1+alpha2}\end{equation}
    \end{subfigure}\hspace{1.5em}
    \begin{subfigure}{0.33\textwidth}
    \begin{equation}\label{eq:hh}\tf{hh}=\tf{id}\end{equation}
    \end{subfigure}

    \hspace{-1.5em}
    \begin{subfigure}{0.53\textwidth}
      \begin{equation}\label{eq:swap}\tf{swap}=\tf{swap-def}\end{equation}
    \end{subfigure}
   
    \vspace{-0.5em}
    \hspace{-1.5em}
    \begin{subfigure}{0.67\textwidth}
      \begin{equation}\label{eq:euler}\tf{euler-left}=\tf{euler-right}\end{equation}
    \end{subfigure}

    \vspace{0.6em}\noindent\rule{\textwidth}{0.4pt}
  
    \vspace{0em}
    \hspace{-1em}
    \begin{subfigure}{0.45\textwidth}
      \begin{equation}\label{eq:conjhcalpha}\tf{conj-h-calpha-right}=\tf{conj-h-calpha-left}\end{equation}
      \end{subfigure}\hspace{2em}
      \begin{subfigure}{0.45\textwidth}
        \begin{equation}\label{eq:conjcalphah}\tf{conj-calpha-h-right}=\tf{conj-calpha-h-left}\end{equation}
      \end{subfigure}
    
    \vspace{0.3em}
    \hspace{-1em}
    \begin{subfigure}{0.45\textwidth}
      \begin{equation}\label{eq:conjhccpi}
        \tf{conj-h-cccpi-right}=\tf{conj-h-cccpi-left}
      \end{equation}
    \end{subfigure}\hspace{2em}
    \begin{subfigure}{0.45\textwidth}
      \begin{equation}\label{eq:conjccpih}
        \tf{conj-ccpi-ch-right}=\tf{conj-ccpi-ch-left}\end{equation}
    \end{subfigure}
    \vspace{0.2em}
  \end{center}\end{minipage}}
  \caption{\label{fig:cqcaxioms}\normalfont Relations $\et{R}_{\textup{c}}$ of $\cat{CQC}$.}
\end{figure*}

We show in the following that the conjugation law can actually be derived for any \cat{CQC} circuits. To do so, we need two ingredients. The first one consists in equipping $\cat{CQC}$ with the dagger functor $(\cdot)^\dagger$ defined as follows, and which provides a natural way to associate an inverse to any quantum circuit.
\begin{gather*}
  (\tf{h})^\dagger\!=\!\tf{h} \hspace{3em}
  (\tf{alpha})^\dagger\!=\!\tf{minusalpha} \hspace{3em}
  (\ctrl (C))^\dagger\!=\!\ctrl (C^\dagger)
\end{gather*}

\begin{proposition}\label{prop:daggeridentity}
  ${\cat{CQC}\vdash C^\dagger\circ C=\textup{id}_n=C\circ C^\dagger}$ for any ${C\in\cat{CQC}(n,n)}$.
\end{proposition}
\begin{proof}
  By induction on $C$ where the base case ${C=\tf{h}}$ is directly proved by \cref{eq:hh} and the base case ${C=\tf{alpha}}$ is proved by $\tf{alpha}\tf{minusalpha}=\tf{0}=\tf{0}\tf{2pi}=\tf{2pi}=\textup{id}_0$ using \cref{eq:2pi,eq:addition}. \qed
\end{proof}

The second ingredient to prove the conjugation law is the ability to provably reduce the number of controls of a gate. To do so, we derive in \cref{app:qcconjugationproof} the following equations that intuitively reduce the number of controls of a gate at the cost of introducing other gates with fewer controls. This is particularly useful for establishing the correspondence with the vanilla quantum circuits. Indeed, we can transform any controllable quantum circuit into one that contains only gates controlled by up to two qubits, and all such gates correspond to a certain vanilla quantum circuit.
\begin{equation}\label{eq:chreduce}
    \smalltf{chaltdef_01}=\smalltf{chaltdef_11}
\end{equation}
\begin{equation}\label{eq:ccalphareduce}
    \smalltf{ccalphadef_01}=\smalltf{ccalphadef_10}
\end{equation}
\begin{equation}\label{eq:cccpireduce}
    \smalltf{cccpidef_01}=\smalltf{cccpidef_10}
\end{equation}

\begin{proposition}\label{prop:qcconjugation}
  $\cat{CQC}$ is a conjugated PROP.
\end{proposition}
\begin{proof}
  \cref{app:conjugationcondition} gives a sufficient condition for the conjugation law to hold in the case of a unitary controlled PROP, such as $\cat{CQC}$ (see \cref{def:untarycontrolledprop}). The idea is to transform any controllable quantum circuit $C$ into an equivalent one $C'$ that contains only the gates $\raisebox{0.5pt}{\protect\tf{alpha}}$, $\ctrl(\!\raisebox{0.5pt}{\protect\tf{alpha}}\!)$, $\ctrl(\ctrl(\!\raisebox{0.5pt}{\protect\tf{pi}}\!))$ and $\tf{h}$. This is achieved using \cref{eq:chreduce,eq:ccalphareduce,eq:cccpireduce} and their controlled versions. Then the sufficient condition states that it is enough to prove the conjugation law for individual gates. The details are given in \cref{app:qcconjugationproof}.\qed
\end{proof}

The conjugation law is very powerful and can be used to prove many equations. In particular, we can generalize \cref{eq:ccalphareduce,eq:cccpireduce} by defining the circuit ${\mu^n(\alpha)\in\cat{CQC}(n,n)}$ for any ${\alpha\in\R}$ and ${n\in\N}$ as follows.
\begin{gather*}\label{eq:mudef}
  \hspace{0.5em}\mu^0(\alpha)\!\defeq\! \tf{alpha} \hspace{3em} 
  \mu^1(\alpha)\!\defeq\!\ctrl(\!\tf{alpha}\!) \\[0.5em]
  \smalltf{mcalpha_ctrl}\!\defeq\!\smalltf{mcalpha-induc-def_ctrl}
\end{gather*}

Intuitively, $\mu^n(\alpha)$ is similar to $\lambda^n(\alpha)$ and implements a Z-rotation gate with parameter $\alpha$ controlled by the first ${n-1}$ qubits. It is provably equivalent to directly controlling an \tf{alpha} gate.

\begin{proposition}\label{prop:ctrlunfolding}
  ${\cat{CQC}\vdash \ctrl^n (\!\raisebox{0.5pt}{\protect\tf{alpha}}\!)=\mu^n (\alpha)}$ for any $n\in\N$.
\end{proposition}
\begin{proof}
  By induction on $n$. The base cases ${n=0}$ and ${n=1}$ are trivial and the case ${n=2}$ is proved in \cref{app:qcconjugationproof} as \cref{eq:ccalphareduce}. The induction case ${n+2}$ with ${n>0}$ is proved using the conjugation law (\cref{prop:qcconjugation}). The details are given in \cref{app:cqccompleteness}.\qed
\end{proof}

We are now ready to prove completeness for controllable quantum circuits by reducing it to the completeness of vanilla quantum circuits.

\begin{theorem}\label{th:cqccompleteness}
	The controlled PROP $\cat{CQC}$ is universally complete for $\cat{Qubit}$, meaning that the functor ${\interp{\cdot}_{\textup{c}}:\cat{CQC}\to\cat{Qubit}}$ is full and faithful.
	\end{theorem}
\begin{proof}
  Universality is already known. The completeness of $\cat{CQC}$ is reduced to the completeness of $\cat{QC}$ (see \cref{th:qccompleteness}). To do so, we define an encoding map ${E:\cat{CQC}\to\cat{QC}}$ and a decoding map ${D:\cat{QC}\to\cat{CQC}}$ to transform a proof in $\cat{QC}$ into a proof in $\cat{CQC}$. The details are given in \cref{app:cqccompleteness}. \qed
\end{proof}

It is remarkable that $\et{R}_{\textup{c}}$ only contains relations acting on at most three qubits, whereas this was not possible in the PROP formalism. This is mainly possible thanks to the context rule saying that if we have ${\cat{CQC}\vdash C_1=C_2}$ then ${\cat{CQC}\vdash \ctrl(C_1)=\ctrl(C_2)}$ comes for free, by definition of the control as a constructor. In some sense, the controlled PROP formalism extracts the structural equations that are specific to control. 

As \cref{eq:conjhcalpha,eq:conjcalphah,eq:conjhccpi,eq:conjccpih} are instances of the conjugation law, one can get rid of these equations by directly considering controllable quantum circuits as a conjugated PROP.
\begin{corollary}\label{cor:ccqc}
  \begingroup\crefrangeformat{equation}{(#3#1#4) to (#5#2#6)}
  The conjugated PROP $\cat{CCQC}$ generated by \tf{alpha} and \tf{h} and the relations \cref{eq:2pi,eq:addition,eq:hh,eq:swap,eq:euler} is universally complete.
  \endgroup
\end{corollary}

This means that \cref{eq:2pi,eq:addition,eq:hh,eq:swap,eq:euler} are sufficient to have a complete graphical language of quantum circuits defined as a conjugated PROP. Moreover, we show that all these relations are necessary, meaning that none can be removed without losing completeness. This is proved in \cref{prop:minimality} in \cref{app:cqccompleteness}.


\section{Controllable quantum circuits with ancillae}\label{sec:aqc}

In this section, we introduce the controllable quantum circuits with auxiliary qubits, a.k.a. ancillae. This provides an example of a controlled PROP that does not consist solely of endomorphisms and demonstrates that, similarly to the vanilla quantum circuit case  \cite{extensions}, ancillae can be embedded into controllable quantum circuits.

\begin{definition}[controllable quantum circuits with ancillae]
    Let $\cat{AQC}$ be the controlled PROP of \emph{controllable quantum circuits with ancillae} generated by the following generators (where ${\alpha\in\R}$) and the relations ${\et{R}_{\textup{c}}\cup\et{R}_{\textup{a}}}$ depicted in \cref{fig:cqcaxioms} and \cref{fig:aqcaxioms}.
  \vspace{-0.5em}
  \begin{gather*}
    \!\tf{alpha}\!:0\to0 \hspace{3em}
    \tf{h}:1\to 1 \hspace{3em}
    \tf{winit}:0\to1 \hspace{3em}
    \tf{wterm}:1\to0
  \end{gather*}
\end{definition}

The additional equations of \cref{fig:aqcaxioms} can be interpreted as follows: $\tf{winit}$ is a point that acts as an annihilator of the control in two particular instances (\cref{eq:winitcalpha,eq:winitccpi}), moreover this point has a left inverse (\cref{eq:winitwterm}), and, as a consequence, is subject to the conjugation law (\cref{eq:conjugationancilla}).

\begin{definition}[interpretation]
  Let ${\interp{\cdot}_{\textup{a}}:\cat{AQC}\to\cat{FdHilb}_2}$ be the interpretation of controllable quantum circuits with ancillae inductively defined as the following identity-on-object PROP functor.
  \vspace{-0.5em}
  \begin{gather*}
    \interp{\tf{alpha}}_{\textup{a}}\defeq e^{i\alpha} \hspace{4em}
    \interp{\tf{winit}}_{\textup{a}}\defeq \ket{0} \hspace{4em}
    \interp{\tf{wterm}}_{\textup{a}}\defeq \bra{0} \\[0.5em]
    \interp{\tf{h}}_{\textup{a}}\defeq \ketbra{+}{0}+\ketbra{-}{1} \hspace{4em}
    \interp{\ctrl (C)}_{\textup{a}}\defeq \ketbra{0}{0}\otimes I_{2^n}+\ketbra{1}{1}\otimes \interp{C}_{\textup{a}}
  \end{gather*}
\end{definition}

The two new generators $\tf{winit}$ and $\tf{wterm}$ correspond respectively to a qubit initialization and a post-selected measurement, hence the semantics of an arbitrary \cat{AQC} circuit is a contraction. It is standard to consider quantum circuits with \emph{clean} ancillae, i.e. a circuit where all $\tf{wterm}$ are performed on qubits whose states have been returned to $\ket 0$, to be safely removed without disturbing the state of the other qubits. Such clean releases are achieved if and only if the semantics of the overall circuit is an isometry\footnote{Indeed, a post-selected evolution being an isometry means that all the performed measurements were actually deterministic, hence whenever $\tf{wterm}$ has been applied, the corresponding qubit was already in the state $\ket 0$.}. Since there is an inclusion ${\cat{Iso}\hookrightarrow\cat{FdHilb}_2}$, we can define $\cat{AQC}_\textup{clean}$ and its interpretation ${\interp{\cdot}_{\textup{i}}:\cat{AQC}_\textup{clean}\to\cat{Iso}}$ as the following pullback, which states that $\cat{AQC}_\textup{clean}$ is the restriction of $\cat{AQC}$ to circuits having isometries as interpretation.
\begin{equation*}
  \bigtf{pullback}
\end{equation*}

\begin{figure*}[t]
  \fbox{\begin{minipage}{0.985\textwidth}\begin{center}
    \vspace{-0.8em}
    \hspace{-1.5em}
    \begin{subfigure}{0.24\textwidth}
      \begin{equation}\label{eq:winitwterm}\tf{winitwterm}=\tf{empty}\end{equation}
    \end{subfigure}\hspace{2em}
    \begin{subfigure}{0.29\textwidth}
      \begin{equation}\label{eq:winitcalpha}\tf{winitcalpha}=\tf{winitlong}\end{equation}
    \end{subfigure}\hspace{2em}
    \begin{subfigure}{0.29\textwidth}
      \begin{equation}\label{eq:winitccpi}\tf{winitccpi}=\tf{winitid}\end{equation}
    \end{subfigure}

    \vspace{-0.8em}
    \hspace{-1.5em}
    \begin{subfigure}{0.47\textwidth}
      \begin{equation}\label{eq:conjugationancilla}\tf{conjugationancilla_left}=\tf{conjugationancilla_right}\end{equation}
    \end{subfigure}
    \vspace{0.2em}
  \end{center}\end{minipage}}
  \caption{\label{fig:aqcaxioms}\normalfont Relations $\et{R}_{\textup{a}}$ of $\cat{AQC}$.}
\end{figure*}

We also add to the above commutative diagram the inclusion of $\cat{CQC}$ in $\cat{AQC}_\textup{clean}$, and we may write ${C\in\cat{CQC}(n,n)}$ when ${C\in\cat{AQC}(n,n)}$ contains no \tf{winit} or \tf{wterm} gates. Moreover, we define some useful notations.
\begin{equation*}
    \tf{not}\defeq\tf{notdef} \hspace{4em} 
    \tf{binit}\defeq\tf{binitdef} \hspace{4em} 
    \tf{bterm}\defeq\tf{btermdef}
\end{equation*}

\begin{proposition}\label{prop:pulloutancilla}
    For any ${C\in\cat{AQC}_\textup{clean}(k,k+n)}$ there exists $C'\in\cat{CQC}(m+n+k,m+n+k)$ such that the following equation is derivable.
    \vspace{-0.5em}
    \begin{equation*}
        \cat{AQC}\vdash\tf{ckn}=\tf{cprimemnk}
    \end{equation*}
\end{proposition}
\begin{proof}
    We can always bend the wires to put all \tf{winit} and \tf{wterm} gates on top of the circuit. Moreover, every \tf{winit} and \tf{wterm} that are inside a control can be taken out using \cref{eq:conjugationancilla} (only endomorphisms can be controlled, which implies that there are always as many \tf{winit} as there are \tf{wterm} inside a control). \qed
\end{proof}

\begin{proposition}\label{prop:aqcpoints}
    \tf{winit} and \tf{binit} are points in $\cat{AQC}_\textup{clean}$, meaning that the following equations are derivable for any $C\in\cat{AQC}_\textup{clean}(n,n)$.
    \begin{equation*}
        \cat{AQC}\vdash\tf{winitcc}=\tf{winitidid}
        \hspace{5em}
        \cat{AQC}\vdash\tf{binitcc}=\tf{binitc}
    \end{equation*}
\end{proposition}
\begin{proof}
  By induction on $C$ and using \cref{prop:pulloutancilla}. The detailed proof is given in \cref{app:aqcpoints}.\qed
\end{proof}

\begin{proposition}\label{prop:isoidentity}
    Given ${C\in\cat{CQC}(n+k,n+k)}$, the following equation is derivable whenever it is sound with respect to $\interp{\cdot}_{\textup{i}}$.
    \begin{equation*}
        \cat{AQC}\vdash\tf{initckn}=\tf{initkn}
    \end{equation*}
\end{proposition}
\begin{proof}
  The proof uses a variant of the cosine-sine decomposition first introduced in \cite{extensions}. The detailed proof is given in \cref{app:isoidentity}.\qed
\end{proof}

\begin{theorem}
    The controlled PROP $\cat{AQC}_\textup{clean}$ is universally complete for $\cat{Iso}$, meaning that the functor $\interp{\cdot}_{\textup{i}}:\cat{AQC}_\textup{clean}\to\cat{Iso}$ is full and faithful.
\end{theorem}
\begin{proof}
  Universality is already known; we only need to prove completeness. To do so, let ${C_1,C_2\in\cat{AQC}_{\textup{clean}}(k,n+k)}$ be such that ${\interp{C_1}_{\textup{i}}=\interp{C_2}_{\textup{i}}}$. Applying \cref{prop:pulloutancilla} to $C_i$ gives the circuit ${C_i'\in\cat{CQC}(m_i+n+k,m_i+n+k)}$. Assume w.l.o.g. that $m_1\ge m_2$. Then, the following derivation proves completeness, where the four wires depict in reality $m_1{-}m_2$, $m_2$, $n$ and $k$ wires respectively.
    \begin{multline*}
        \cat{AQC}\vdash C_1
        \eqcref{prop:pulloutancilla}\tf{iqccompleteness_01}
        \eqcref{prop:daggeridentity}\tf{iqccompleteness_02}\\[0.4em]
        \eqcref{prop:isoidentity}\tf{iqccompleteness_03}
        \eqoneref{eq:winitwterm}\tf{iqccompleteness_04}
        \eqcref{prop:pulloutancilla}C_2
    \end{multline*}
    where \cref{prop:isoidentity} is applicable because the equality
    \begin{equation*}
        \interp{C_1}_{\textup{i}}=\biginterp{\tf{iqccompleteness_sem_01_alt}}_{\textup{i}} = \biginterp{\tf{iqccompleteness_sem_02_alt}}_{\textup{i}}=\interp{C_2}_{\textup{i}}
    \end{equation*}
    implies the following equality.
    \begin{equation*}
        \biginterp{\tf{iqccompleteness_sem_03_alt}}_{\textup{i}} = \biginterp{\tf{iqccompleteness_sem_04_alt}}_{\textup{i}}
    \end{equation*}\qed
\end{proof}

Similarly to the ancilla-free case, one can consider quantum circuits with ancillae as a conjugated PROP and get rid of the equations that are instances of the conjugation law.

\begin{corollary}
  \begingroup\crefrangeformat{equation}{(#3#1#4) to (#5#2#6)}
  Let $\cat{CAQC}$ be the conjugated PROP generated by \tf{alpha}, \tf{h}, \tf{winit} and \tf{wterm} and the relations \cref{eq:2pi,eq:addition,eq:hh,eq:swap,eq:euler} together with \cref{eq:winitwterm,eq:winitcalpha,eq:winitccpi}. Then, its restriction to isometric circuits $\cat{CAQC}_\textup{clean}$ is universally complete.
  \endgroup
\end{corollary}


\section{Polycontrolled PROP}\label{sec:polycontrol}

Throughout the paper, we focused on PROPs that are equipped with a single control functor. It is natural to consider \emph{polycontrolled PROPs}, i.e.~PROPs that have multiple control functors.

Considering polycontrolled PROPs instead of controlled PROPs can sometimes simplify diagrammatic reasoning. For instance, while controllable quantum circuits are defined using two generators and one control functor (see \cref{sec:cqc}), we can also define them using one generator and two control functors. This idea has first been explained in \cite{heunen25}, and can be expressed in the controlled PROP formalism as follows: if we only have global phases \tf{alpha} for any $\alpha\in\R$, and two control functors $\ctrl_Z$ and $\ctrl_X$, we can still get a universal graphical language for $\cat{Qubit}$ with the following interpretation.
\begin{gather*}
    \interp{\tf{alpha}}\defeq e^{i\alpha} \\[0.2em]
    \interp{\ctrl_Z(C)}\defeq \ketbra{0}{0}\otimes I_{2^n}+\ketbra{1}{1}\otimes \interp{C} \\[0.2em]
    \interp{\ctrl_X(C)}\defeq \ketbra{+}{+}\otimes I_{2^n}+\ketbra{-}{-}\otimes \interp{C}
\end{gather*}

Intuitively $\ctrl_Z$ allows implementing Z-rotations while $\ctrl_X$ allows implementing X-rotations. Moreover, Z- and X-rotations together with their multi-controlled version can express any unitary. For instance, the Hadamard gate can be implemented by $\tf{minuspiover4}\otimes\ctrl_Z(\!\raisebox{0.5pt}{\protect\tf{piover2}}\!)\circ\ctrl_X(\!\raisebox{0.5pt}{\protect\tf{piover2}}\!)\circ\ctrl_Z(\!\raisebox{0.5pt}{\protect\tf{piover2}}\!)$.

In the following, we define some notable properties that multiple coexisting control functors may satisfy.

\begin{definition}[compatibility]\label{def:compatibility}
    Given a polycontrolled PROP $\cat{P}$, we say that its control functors $\ctrl_1$ and $\ctrl_2$ are \emph{compatible} if the following equation is satisfied for any $f\in\cat{P}(n,n)$.
    \begin{equation}\label{eq:compatibility}
        \hspace{-1.1em}\begin{array}{ccc}
            \ctrl_1(\ctrl_2 (f))\circ(\sigma_{1,1}\otimes \textup{id}_{n})&=&(\sigma_{1,1}\otimes \textup{id}_{n})\circ\ctrl_2(\ctrl_1 (f))\\[.4em]
            \tf{compatibility_left}&=&\tf{compatibility_right}
        \end{array}
    \end{equation}
\end{definition}

\begin{example}
    In $\cat{FdHilb}_2$, $\ctrl_{\ket{1}}$ is compatible with $\ctrl_{\ket{0}}$ because each control can be obtained by conjugating the control wire of the other control with a NOT gate that slides through the swaps.
    However, $\ctrl_{\ket{1}}$ is not compatible with the control functor $\ctrl_\sharp:U\mapsto I_2\otimes X_nUX_n$ where $U$ acts on $n$ qubits and where $X_n$ applies NOT gates to $n$ qubits in parallel. Intuitively, $\ctrl_\sharp$ is independent of the state of the control qubit and always conjugates the target by a series of NOT gates in parallel.
\end{example}

\begin{definition}[commutativity]\label{def:commutativity}
    Given a polycontrolled PROP $\cat{P}$, we say that its control functors $\ctrl_1$ and $\ctrl_2$ \emph{commute} if the following equation is satisfied for any $f,g\in\cat{P}(n,n)$.
    \begin{equation}\label{eq:commutativity}
        \begin{array}{ccc}
            \ctrl_2(g)\circ \ctrl_1(f) &=& \ctrl_1(f)\circ \ctrl_2(g)\\[.4em]
            \tf{commutativity_left}&=&\tf{commutativity_right}
        \end{array}
    \end{equation}
\end{definition}

\begin{example}
    In $\cat{FdHilb}_2$, $\ctrl_{\ket{1}}$ commutes with $\ctrl_{\ket{0}}$ whereas it does not with $\ctrl_{\ket{-}}$.
\end{example}

\begin{definition}[exhaustivity]\label{def:exhaustivity}
    Given a polycontrolled PROP $\cat{P}$ with a family of $\ell$ commuting control functors ${(\ctrl_k)_{k\in[\ell]}}$, we say that the family is \emph{exhaustive} if the following equation is satisfied for any $f\in\cat{P}(n,n)$.
    \begin{equation}\label{eq:exhausivity}
        \begin{array}{ccc}
            \ctrl_\ell(f)\circ \ldots \circ \ctrl_1(f) &=& \textup{id}_1\otimes f\\[.4em]
            \tf{exhaustivity_left}&=&\tf{exhaustivity_right}
        \end{array}
    \end{equation}
\end{definition}

\begin{example}\label{ex:exhaust}
    In $\cat{FdHilb}_d$, the family of control functors ${(\ctrl_{\ket{k}})_{0\le k<d}}$ is exhaustive. Intuitively, each controlled operation ${\ctrl_{\ket{k}}(f)}$ fires $f$ if the control qubit is in the state $\ket{k}$. Thus, if we apply all these controlled operations sequentially, then $f$ is always fired exactly once.
\end{example}

\section{Conclusion}\label{sec:conclusion}

In this paper, we introduced controlled PROPs as an extension of the standard PROP formalism enabling diagrammatic reasoning with control as a constructor. Beyond quantum computing, we expect controlled PROPs to be useful in other domains where conditional behavior plays a central role, such as classical reversible computing. Note that the notion of control functor can be straightforwardly generalized to arbitrary symmetric monoidal categories --~as explained in \cref{app:controlsmc}~-- and also to weaker structures like braided monoidal categories. 

We considered, in this paper, control as a constructor, independently of its actual physical implementation. We leave a full development of this question for future investigation; however we mention a no-go theorem concerning the possibility of controlling unknown operations using standard quantum circuit formalism \cite{araujo2014}. As pointed out by the authors of this result, the no-go theorem does not apply in practice, as any physical implementation of an unknown unitary provides additional information that makes control possible. Moreover, there is evidence that such control can be implemented in practice \cite{friis2014,zhou2011}. The authors of the no-go theorem even argue that the quantum circuit formalism should be extended to capture the possibility of controlling unknown unitaries, which is a feature of the controlled PROP of controllable quantum circuits we introduced. Furthermore, in the present paper, this no-go theorem does not apply, as we work in a white-box setting, i.e. we assume that the implementation of the unitary as a circuit is known.

Finally, another direction for future research is to consider controllable quantum circuits for quantum circuit optimization tasks, since they provide simpler rewriting rules than those of vanilla quantum circuits.

\section*{Acknowledgements}

The authors want to thank Alexandre Clément, Emmanuel Jeandel, Louis Lemonnier, William Schober, and Scott Wesley for fruitful discussions. This work is supported by the Plan France 2030 through the PEPR integrated project EPiQ ANR-22-PETQ-0007 and the HQI platform ANR-22-PNCQ-0002; and by the European Union through the MSCA Staff Exchange project Qcomical HORIZON-MSCA-2023-SE-01. The project is also supported by the Maison du Quantique MaQuEst.

\newpage
\bibliographystyle{splncs04}
\bibliography{ref}

\appendix
\renewcommand\theHsection{\thesection}

\section*{Appendix content}
\cref{app:conjugationcondition} gives a sufficient condition for the conjugation law to hold in the context of a unitary controlled PROP. \cref{app:qcconjugationproof} contains a detailed proof of \cref{prop:qcconjugation}. \cref{app:cqccompleteness} contains detailed proofs of \cref{prop:ctrlunfolding} and \cref{th:cqccompleteness}. \cref{app:aqcpoints} contains a detailed proof of \cref{prop:aqcpoints}. \cref{app:csdvariant} proves a variant of the cosine-sine decomposition. \cref{app:isoidentity} contains a detailed proof of \cref{prop:isoidentity}. \cref{app:controlsmc} contains an extension of the definition of control functors for arbitrary symmetric monoidal categories.


\section{Sufficient condition for the conjugation law}\label[appendix]{app:conjugationcondition}

In the following, the notation $[n]$ denotes the set $\{i:i\in\N, 1\le i\le n\}$ containing all integers between $1$ and $n$ (included). In particular, if $n\le0$ then $[n]=\varnothing$.

\begin{definition}[unitary controlled PROP]\label{def:untarycontrolledprop}
    A \emph{unitary controlled PROP} $\cat{P}$ is a controlled PROP of endomorphisms together with a dagger functor $(\cdot)^\dagger$ satisfying $(\ctrl(f))^\dagger=\ctrl(f^\dagger)$ and $\cat{P}\vdash f\circ f^\dagger=\textup{id}_n=f^\dagger\circ f$ for any $f\in\cat{P}(n,n)$.
\end{definition}

\begin{definition}[reducibility]
    Given a unitary controlled PROP $\cat{P}$ and a set $\mathcal{G}\subseteq\cup_{k\in\N}\cat{P}(k,k)$ of endomorphisms, let $\cat{P}\vert_{\mathcal{G}}(n,n)\subseteq\cat{P}(n,n)$ be subclasses of endomorphisms inductively defined below. We say that $\cat{P}$ is \emph{$\mathcal{G}$-reducible} if for any $f\in\cat{P}(n,n)$ there exists $f'\in\cat{P}\vert_{\mathcal{G}}(n,n)$ such that $\cat{P}\vdash f=f'$.
    \begin{gather*}
        \infer{f\in\cat{P}\vert_{\mathcal{G}}(n,n)}{f:n\to n\in\mathcal{G}} \hspace{3em}
        \infer{(g\circ f)\in\cat{P}\vert_{\mathcal{G}}(n,n)}{f,g\in\cat{P}\vert_{\mathcal{G}}(n,n)} \hspace{3em}
        \infer{(f\otimes g)\in\cat{P}\vert_{\mathcal{G}}(n+m,n+m)}{f\in\cat{P}\vert_{\mathcal{G}}(n,n) & g\in\cat{P}\vert_{\mathcal{G}}(m,m)}
    \end{gather*}
\end{definition}

\begin{lemma}\label{lem:conjugationcondition}
    Given a $\mathcal{G}$-reducible unitary controlled PROP $\cat{P}$,
    \begin{equation}\label{eq:conjugationcondition}\tag{$\star$}
        \hspace{-0.5em}\begin{array}{c}
            \forall x:n\to n,y:m\to m\in\mathcal{G} \\
            \forall i\in[n+m-1]
        \end{array},\; \cat{P}\vdash \tf{conjugation_YZ_left} = \tf{conjugation_YZ_right}
    \end{equation}
    implies
    \begin{equation*}
        \begin{array}{c}
            \forall f,g,h\in\cat{P}(n,n) \\
            \suchthat \cat{P}\vdash h\circ g=\textup{id}_n
        \end{array},\; \cat{P}\vdash \tf{conjugation_gfh_left} = \tf{conjugation_gfh_right}
    \end{equation*}
    where $f_{x,i}$ and $g_{y,i}$ are defined as follows.
    \begin{gather*}
        f_{x,i}\defeq\textup{id}_{\max(0,i-n)}\otimes x\otimes\textup{id}_{\max(0,m-i)}
        \hspace{2em}
        g_{y,i}\defeq\textup{id}_{\max(0,n-i)}\otimes y\otimes\textup{id}_{\max(0,i-m)}
    \end{gather*}
\end{lemma}
\begin{proof}
    As $\cat{P}$ is $\mathcal{G}$-reducible, any morphism in $\cat{P}(n,n)$ can be transformed into a morphism in $\cat{P}\vert_{\mathcal{G}}(n,n)$.
    Moreover, notice that the coherence laws of PROPs (\cref{fig:coherenceprop}) ensure that for any $f_0\in\cat{P}\vert_{\mathcal{G}}(n,n)$ there exist $\ell\ge1$ and $p_k,q_k\in\N$, $x_k:n_k\to n_k\in\mathcal{G}$ satisfying $p_k+n_k+q_k=n$ for all $1\le k\le \ell$, such that
    \begin{gather*}
        \cat{P}\vdash f_0=(\textup{id}_{p_{\ell}}\otimes x_{\ell}\otimes\textup{id}_{q_{\ell}})\circ(\textup{id}_{p_{\ell-1}}\otimes x_{\ell-1}\otimes\textup{id}_{q_{\ell-1}})\circ\dots\circ(\textup{id}_{p_1}\otimes x_1\otimes\textup{id}_{q_1})
    \end{gather*}

    Intuitively, we can transform any morphism $f_0$ into a morphism $f_0'$ that is a sequential composition of $\ell$ layers each containing a single morphism of $\mathcal{G}$ sandwiched between identities.
    Applied to $f$ and $g$, this yields $f'$ and $g'$ of the above form such that $\cat{P}\vdash f=f'$ and $\cat{P}\vdash g=g'$. Moreover,
    \begin{equation*}
        \cat{P}\vdash h=h\circ g'\circ (g')^\dagger=h\circ g\circ (g')^\dagger=(g')^\dagger
    \end{equation*}

    Then, assuming \eqref{eq:conjugationcondition}, we need to prove the following equation.
    \begin{equation}\label{eq:conjugationprime}
        \cat{P}\vdash\tf{conjugation_big_prime_left}=\tf{conjugation_big_prime_right}
    \end{equation}

    Let $\ell_f$ and $\ell_g$ be the number of layers of $f'$ and $g'$ respectively. In the rest of the proof, we prove \cref{eq:conjugationprime} by induction on $(\ell_g,\ell_f)$ with lexicographic order.
    If $f'=f_2'\circ f_1'$ where $\ell_{f_i}$ is the number of layers of $f_i'$, then
    \begin{gather*}
        \tf{conjugation_big_prime_left}
        =\tf{conjproof_f_01} \\[0.5em]
        =\tf{conjproof_f_02} 
        =\tf{conjproof_f_03} \\[0.5em]
        \overset{\textup{IH}}{=}\tf{conjproof_f_04} \\[0.5em]
        =\tf{conjproof_f_05} 
        =\tf{conjproof_f_06}
        =\tf{conjugation_big_prime_right}
    \end{gather*}
    where the induction hypothesis is applied for $(\ell_g,\ell_{f_i})<(\ell_g,\ell_{f})$. If $g'=g_2'\circ g_1'$ where $\ell_{g_i}$ is the number of layers of $g_i'$, then
    \begin{gather*}
        \tf{conjugation_big_prime_left}
        =\tf{conjproof_g_01} \\[0.5em]
        \overset{\textup{IH}}{=}\tf{conjproof_g_02} \\[0.5em]
        \overset{\textup{IH}}{=}\tf{conjproof_g_03}
        =\tf{conjugation_big_prime_right}
    \end{gather*}
    where the first induction hypothesis is applied for $(\ell_{g_1},\ell_{f}+2\ell_{g_2})<(\ell_g,\ell_{f})$ and the second induction hypothesis is applied for $(\ell_{g_2},\ell_{f})<(\ell_g,\ell_{f})$.

    The base cases $\ell_f=\ell_g=1$ are the cases where $f'$ and $g'$ have a single layer. Let 
    \begin{equation*}
        f'=(\textup{id}_{p}\otimes x\otimes\textup{id}_{q})
        \hspace{3em}\text{and}\hspace{3em}
        g'=(\textup{id}_{r}\otimes y\otimes\textup{id}_{s})
    \end{equation*}
    where $p,q,r,s\in\N$ and $x:n\to n,y:m\to m\in\mathcal{G}$. If $p\ge r+m$ or $r\ge p+n$ then $f'$ and $g'$ trivially commute (as well as $(\textup{id}_1\otimes g')$ and $\ctrl(f')$), and so the statement is straightforwardly proved using the fact that $\cat{P}$ is unitary. Otherwise, there exists a unique $i\in[n+m-1]$ such that
    \begin{equation*}
        \cat{P}\vdash f'=\textup{id}_{\min(p,r)}\otimes f_{x,i}\otimes\textup{id}_{\min(q,s)}
        \hspace{1em}\text{and}\hspace{1em}
        \cat{P}\vdash g'=\textup{id}_{\min(p,r)}\otimes g_{y,i}\otimes\textup{id}_{\min(q,s)}
    \end{equation*}

    Hence, we prove the base cases as follows.
    \begin{gather*}
        \tf{conjproof_base_01}
        =\tf{conjproof_base_02}
        =\tf{conjproof_base_03} \\[0.5em]
        \eqoneref{eq:conjugationcondition}\tf{conjproof_base_04} 
        =\tf{conjproof_base_05}
        =\tf{conjproof_base_06}
    \end{gather*}\qed
\end{proof}


\section{Proof of \cref{prop:qcconjugation}}\label[appendix]{app:qcconjugationproof}

The idea is to apply \cref{lem:conjugationcondition} to $\cat{CQC}$. To do so, we prove that the sufficient condition \eqref{eq:conjugationcondition} is satisfied in $\cat{CQC}$. The following equations are used as intermediate results and are proved at the end of this section.

\begin{equation}\label{eq:0}
    \tf{0_01}=\tf{0_04}
\end{equation}

\begin{equation}\label{eq:minuspi}
    \tf{minuspi_01}=\tf{minuspi_03}
\end{equation}

\begin{equation}\label{eq:heuler}
    \tf{heuler_01}=\tf{heuler_03}
\end{equation}

\begin{equation}\label{eq:xcalpha}
    \tf{xcalpha_01}=\tf{xcalpha_06}
\end{equation}

\begin{equation}\label{eq:heulervar}
    \tf{heulervar_01}=\tf{heulervar_05}
\end{equation}

\begin{equation}\label{eq:conjhh}
    \tf{conjhh_04}=\tf{conjhh_01}
\end{equation}

\begin{equation}\label{eq:conjhccpi_var}
    \tf{conjhccpi_04}=\tf{conjhccpi_01}
\end{equation}

\begin{equation}\label{eq:conjccpih_var}
    \tf{conjccpih_04}=\tf{conjccpih_01}
\end{equation}

\begin{lemma}\label{lem:cqcreducing}
    $\cat{CQC}$ is $\mathcal{G}$-reducible where $\mathcal{G}$ is defined as follows.
    \begin{gather*}
        \mathcal{G}\defeq
        \left\{\textup{id}_1\right\}
        \cup\left\{\tf{alpha}:\alpha\in\R\right\}
        \cup\left\{\ctrl(\!\tf{alpha}\!):\alpha\in\R\right\}
        \cup\left\{\ctrl(\ctrl(\!\tf{pi}\!))\right\}
        \cup\left\{\tf{h}\right\}
    \end{gather*}
\end{lemma}
\begin{proof}
    Given a morphism $f\in\cat{CQC}(n,n)$, we explain how to transform $f$ into a morphism $f'\in\cat{CQC}\vert_{\mathcal{G}}(n,n)$. In the following, what we call a gate is either $\tf{alpha}$ or $\tf{h}$.
    Notice that all $\textup{id}_0$ can be removed by applying \cref{eq:paridentity,eq:seqidentity}. Also, all $\sigma_{1,1}$  can be removed using \cref{eq:swap}.
    Moreover, as $\ctrl$ is a functor and $\cat{CQC}$ contains only endomorphisms, we can always apply $\ctrl(g\circ f)=\ctrl(g)\circ\ctrl(f)$ to split each control into multiple controlled gates.
    Then, we iteratively reduce the number of controls of the remaining gates using \cref{eq:chreduce,,eq:ccalphareduce,,eq:cccpireduce} as well as their controlled versions (where \cref{eq:ccalphareduce} is only applied when $\alpha\ne\pi$ to avoid infinite loops). The coherence laws ensure that this process can be applied to gates that are controlled by non-adjacent wires, and we can remove the implicit $\sigma_{1,1}$ gates that are hidden using \cref{eq:swap}. At the end of this process,  we get a morphism $f'\in\cat{CQC}\vert_{\mathcal{G}}(n,n)$.\qed
\end{proof}

\begin{proof}[\cref{prop:qcconjugation}]
    \cref{prop:daggeridentity,lem:cqcreducing} ensure that we can apply \cref{lem:conjugationcondition} to $\cat{CQC}$. Proving that \eqref{eq:conjugationcondition} holds amounts to proving that for any $x:n\to n,y:m\to m\in\mathcal{G}$ and $i\in[n+m-1]$,
    \begin{equation*}
        \cat{CQC}\vdash\tf{conjugation_YZ_left}=\tf{conjugation_YZ_right}
    \end{equation*}

    We give below a table summing up all the cases. Fortunately, this is trivial for most cases. In particular, the fact that $\ctrl(\textup{id}_n)=\textup{id}_{1+n}$ and \cref{prop:daggeridentity} can straightforwardly be used when $x=\textup{id}_1$ or $y=\textup{id}_1$ or when $x=\tf{alpha}$ or $y=\tf{beta}$ for some $\alpha,\beta\in\R$ (this remark is referred to as $\heartsuit$ in the table).
    Moreover, all circuits containing a single controlled \tf{alpha} gate provably commute. Here is an example.
    \begin{gather*}
        \tf{ctrlphasecommute_exemple_01}
        =\tf{ctrlphasecommute_exemple_03}
        =\tf{ctrlphasecommute_exemple_04}
        =\tf{ctrlphasecommute_exemple_06}
    \end{gather*}

    Hence, all the base cases where $x$ and $y$ are either $\ctrl(\!\raisebox{1pt}{\tf{alpha}}\!)$ or $\ctrl(\ctrl(\!\raisebox{1pt}{\tf{pi}}\!))$ can be proved easily (this remark is referred to as $\clubsuit$ in the table). Here is an example for $y=\ctrl(\ctrl(\!\raisebox{1pt}{\tf{pi}}\!))$ and $x=\ctrl(\!\raisebox{1pt}{\tf{alpha}}\!)$.
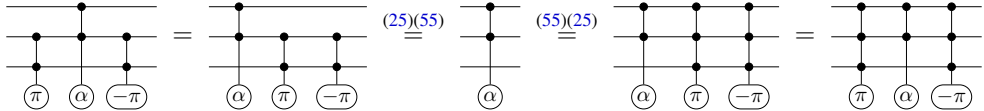
    \begin{gather*}
        \tf{conjphase_example_01}
        =\tf{conjphase_example_02}
        \eqtworef{eq:addition}{eq:0}\tf{conjphase_example_03} 
        \eqtworef{eq:0}{eq:addition}\tf{conjphase_example_04}\\[0.5em]
        =\tf{conjphase_example_05}
    \end{gather*}

    Finally, the remaining cases are either relations in $\et{R}_{\textup{c}}$ or provable in $\cat{CQC}$. Here is the table summing up all cases. The proof of \cref{eq:conjccpih_var,,eq:conjhccpi_var,,eq:conjhh} can be found at the end of this section.

    {\renewcommand{\arraystretch}{1.5}
    \begin{center}
        \begin{tabular}{c|c|c|c|c|c} 
        & $x=\textup{id}_1$ & $x=\tf{alpha}$ & $x=\ctrl(\!\tf{alpha}\!)$ & $x=\ctrl(\ctrl(\!\tf{pi}\!))$ & $x=\tf{h}$ \\\hline
        $y=\textup{id}_1$ & 
        $\heartsuit$ &
        $\heartsuit$ &
        $\heartsuit$ &
        $\heartsuit$ &
        $\heartsuit$ \\\hline
        $y=\tf{beta}$ &
        $\heartsuit$ &
        $\heartsuit$ &
        $\heartsuit$ &
        $\heartsuit$ &
        $\heartsuit$ \\\hline
        $y=\ctrl(\!\tf{beta}\!)$ &
        $\heartsuit$ &
        $\heartsuit$ &
        $\clubsuit$ &
        $\clubsuit$ &
        \begin{tabular}{cc}$i=1$&\eqref{eq:conjcalphah}\end{tabular} \\\hline
        $y=\ctrl(\ctrl(\!\tf{pi}\!))$ & 
        $\heartsuit$ &
        $\heartsuit$ &
        $\clubsuit$ &
        $\clubsuit$ &
        \begin{tabular}{cc}$i=1$&\eqref{eq:conjccpih_var} \\\hline $i=2$&\eqref{eq:conjccpih}\end{tabular} \\\hline
        $y=\tf{h}$ & 
        $\heartsuit$ &
        $\heartsuit$ &
        \begin{tabular}{cc}$i=1$&\eqref{eq:conjhcalpha}\end{tabular} &
        \begin{tabular}{cc}$i=1$&\eqref{eq:conjhccpi_var} \\\hline $i=2$&\eqref{eq:conjhccpi}\end{tabular} &
        \begin{tabular}{cc}$i=1$&\eqref{eq:conjhh}\end{tabular}
        \end{tabular}
    \end{center}}\qed
\end{proof}

\begin{proof}[\cref{eq:0}]
    \begin{gather*}
        \tf{0_01}
        \eqoneref{eq:2pi}\tf{0_02}
        \eqoneref{eq:addition}\tf{0_03}
        \eqoneref{eq:2pi}\tf{0_04}
    \end{gather*}\qed
\end{proof}

\begin{proof}[\cref{eq:minuspi}]
    \begin{gather*}
        \tf{minuspi_01}
        \eqoneref{eq:2pi}\tf{minuspi_02}
        \eqoneref{eq:addition}\tf{minuspi_03}
    \end{gather*}\qed
\end{proof}

\begin{proof}[\cref{eq:heuler}]
    \begin{gather*}
        \tf{heuler_01}
        \eqtworef{eq:hh}{eq:0}\tf{heuler_02}
        \eqoneref{eq:euler}\tf{heuler_03}
    \end{gather*}\qed
\end{proof}

\begin{proof}[\cref{eq:xcalpha}]
    \begin{gather*}
        \tf{xcalpha_01}
        \eqoneref{eq:hh}\tf{xcalpha_02} \\[0.7em]
        \eqoneref{eq:euler}\tf{xcalpha_03} \\[0.7em]
        \eqtworef{eq:2pi}{eq:addition}\tf{xcalpha_04} \\[0.7em]
        \eqoneref{eq:addition}\tf{xcalpha_05}
        \eqoneref{eq:heuler}\tf{xcalpha_06}
    \end{gather*}\qed
\end{proof}

\begin{proof}[\cref{eq:heulervar}]
    \begin{gather*}
        \tf{heulervar_01}
        \eqoneref{eq:heuler}\tf{heulervar_02}
        \eqoneref{eq:addition}\tf{heulervar_03} \\[0.7em]
        \eqoneref{eq:hh}\tf{heulervar_04} \\[0.7em]
        \eqoneref{eq:xcalpha}\tf{heulervar_05}
    \end{gather*}\qed
\end{proof}

\begin{proof}[\cref{eq:chreduce}]
    \begin{gather*}
        \tf{chaltdef_01}
        \eqoneref{eq:heulervar}\tf{chaltdef_02}\\[0.7em]
        \eqoneref{eq:addition}\tf{chaltdef_03}\\[0.7em]
        \eqoneref{eq:conjcalphah}\tf{chaltdef_04}\\[0.7em]
        \eqoneref{eq:addition}\tf{chaltdef_05}
        \eqoneref{eq:conjhcalpha}\tf{chaltdef_06}\\[0.7em]
        \eqoneref{eq:conjhcalpha}\tf{chaltdef_07}
        \eqoneref{eq:addition}\tf{chaltdef_08}\\[0.7em]
        \eqoneref{eq:conjcalphah}\tf{chaltdef_09}
        \eqoneref{eq:addition}\tf{chaltdef_10}\\[0.7em]
        \eqoneref{eq:conjhcalpha}\tf{chaltdef_11}
    \end{gather*}\qed
\end{proof}

\begin{proof}[\cref{eq:conjhh}]
    \begin{gather*}
        \tf{hch_09}
        \eqoneref{eq:hh}\tf{hch_08}
        \eqoneref{eq:chreduce}\tf{hch_07}\\[0.7em]
        \eqoneref{eq:addition}\tf{hch_06}\\[0.7em]
        \eqthreeref{eq:hh}{eq:0}{eq:addition}\tf{hch_05}\\[0.7em]
        \eqoneref{eq:heulervar}\tf{hch_04}\\[0.7em]
        \eqthreeref{eq:hh}{eq:0}{eq:addition}\tf{hch_03}\\[0.7em]
        \eqoneref{eq:heulervar}\tf{hch_02}
        \eqoneref{eq:chreduce}\tf{hch_01}
    \end{gather*}\qed
\end{proof}

\begin{proof}[\cref{eq:conjhccpi_var}]
    \begin{gather*}
        \tf{conjhccpi_04}
        \eqoneref{eq:controlswap}\tf{conjhccpi_03}
        \eqoneref{eq:conjhccpi}\tf{conjhccpi_02}
        =\tf{conjhccpi_01}
    \end{gather*}\qed
\end{proof}

\begin{proof}[\cref{eq:conjccpih_var}]
    \begin{gather*}
        \tf{conjccpih_04}
        \eqoneref{eq:controlswap}\tf{conjccpih_03}
        \eqoneref{eq:conjccpih}\tf{conjccpih_02}
        =\tf{conjccpih_01}
    \end{gather*}\qed
\end{proof}

\begin{proof}[\cref{eq:ccalphareduce}]
    \begin{gather*}
        \tf{ccalphadef_01}
        \eqoneref{eq:addition}\tf{ccalphadef_02}
        \eqthreeref{eq:hh}{eq:2pi}{eq:addition}\tf{ccalphadef_03}\\[0.7em]
        \eqoneref{eq:xcalpha}\tf{ccalphadef_04}
        =\tf{ccalphadef_05} \\[0.7em]
        =\tf{ccalphadef_06} \\[0.7em]
        \eqoneref{eq:addition}\tf{ccalphadef_07}\\[0.7em]
        \eqoneref{eq:conjcalphah}\tf{ccalphadef_08}\\[0.7em]
        \eqoneref{eq:addition}\tf{ccalphadef_09} 
        \eqoneref{eq:conjhcalpha}\tf{ccalphadef_10}
    \end{gather*}\qed
\end{proof}

\begin{proof}[\cref{eq:cccpireduce}]
    \begin{gather*}
        \tf{cccpidef_01}
        =\tf{cccpidef_02}
        \eqoneref{eq:ccalphareduce}\tf{cccpidef_03}\\[0.7em]
        =\tf{cccpidef_04}
        \eqoneref{eq:conjhccpi_var}\tf{cccpidef_05}\\[0.7em]
        \eqoneref{eq:conjhcalpha}\tf{cccpidef_06}\\[0.7em]
        \eqoneref{eq:addition}\tf{cccpidef_07}\\[0.7em]
        \eqoneref{eq:conjccpih_var}\tf{cccpidef_08}\\[0.7em]
        \eqoneref{eq:addition}\tf{cccpidef_09}
        \eqoneref{eq:conjhcalpha}\tf{cccpidef_10}
    \end{gather*}\qed
\end{proof}


\section{Proofs of \cref{prop:ctrlunfolding} and \cref{th:cqccompleteness}}\label[appendix]{app:cqccompleteness}

In this section, we use the set $\mathcal{G}$ defined in \cref{lem:cqcreducing}.

\begin{proof}[\cref{prop:ctrlunfolding}]
  The base cases where $n=0$ and $n=1$ are trivial and the case $n=2$ is proved in \cref{app:qcconjugationproof} as \cref{eq:ccalphareduce}.
  The induction case $n+2$ with $n>0$ is proved by the following derivation (where the second wire represents $n-1$ wires).
  \begin{gather*}
    \ctrl^{n+2} (\!\tf{alpha}\!)
    =\tf{mcalpha_01}
    =\tf{mcalpha_02}\\[.7em]
    \overset{\textup{IH}}{=}\tf{mcalpha_03}\\[.7em]
    \overset{\textup{IH}}{=}\tf{mcalpha_04}\\[.7em]
    \eqoneref{eq:swapconjugation}\tf{mcalpha_05}\\[.7em]
    \eqoneref{eq:conjugation}\tf{mcalpha_06}\\[.7em]
    \overset{\textup{IH}}{=}\tf{mcalpha_07}
    =\mu^{n+2}(\alpha)
  \end{gather*}\qed
\end{proof}

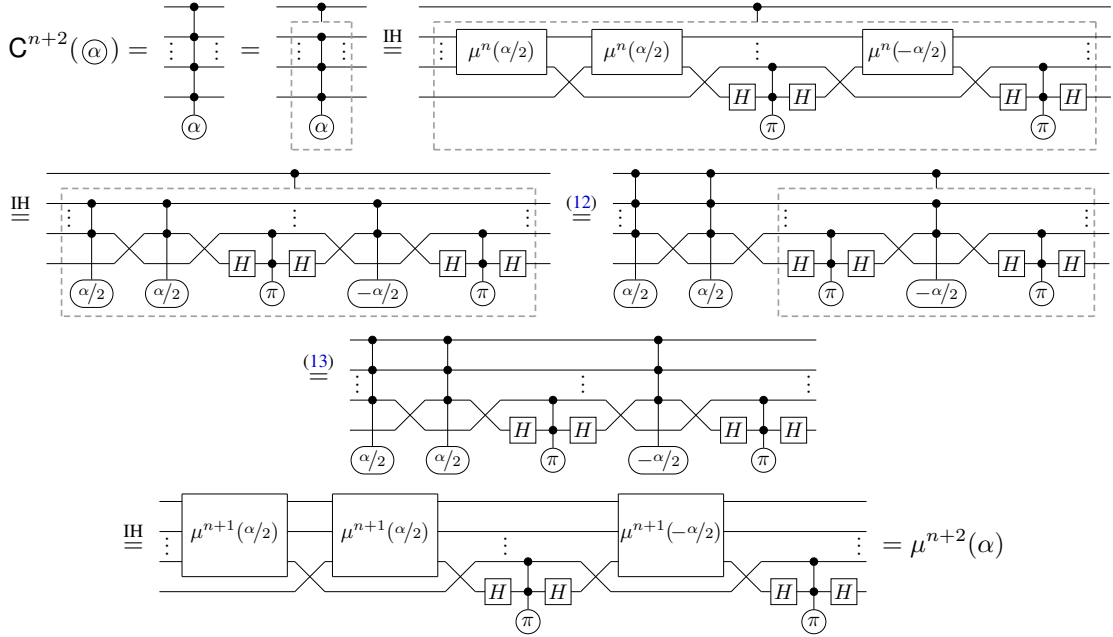
\begin{definition}[encoding]\label{def:encoding}
  Let $E:\cat{CQC}\vert_{\mathcal{G}}\to\cat{QC}$ be the \emph{encoding map} inductively defined as follows.
  \begin{gather*}
    E(C_2\circ C_1) \defeq E(C_2)\circ E(C_1) \hspace{4em}
    E(C_1\otimes C_2) \defeq E(C_1)\otimes E(C_2) \\[0.4em]
    E(\textup{id}_1) \defeq \textup{id}_1 \hspace{4em}
    E(\!\tf{alpha}\!) \defeq \tf{alpha} \hspace{4em}
    E\left(\tf{h}\right) \defeq \tf{h} \\[0.4em]
    E\left(\tf{calphagen}\right) \defeq \tf{zalpha} \hspace{4em}
    E\left(\tf{ccpigen}\right) \defeq \tf{hcnoth}
  \end{gather*}
\end{definition}

\begin{definition}[decoding]\label{def:decoding}
  Let $D:\cat{QC}\to\cat{CQC}\vert_{\mathcal{G}}$ be the \emph{decoding map} inductively defined as follows.
  \begin{gather*}
    D(C_2\circ C_1)\defeq D(C_2)\circ D(C_1) \hspace{4em}
    D(C_1\otimes C_2)\defeq D(C_1)\otimes D(C_2) \\[0.4em]
    D\left(\textup{id}_1\right)\defeq \textup{id}_1 \hspace{4em}
    D\left(\sigma_{1,1}\right)\defeq\tf{swap-def} \\[0.4em]
    D\left(\!\tf{alpha}\!\right)\defeq \tf{alpha} \hspace{4em}
    D\left(\tf{h}\right)\defeq \tf{h} \hspace{4em}
    D\left(\tf{zalpha}\right)\defeq \tf{calpha}\\[0.4em]
    D\left(\tf{cnot}\right)\defeq \tf{hccpih} 
  \end{gather*}
\end{definition}

\begin{lemma}\label{lem:interp}
  $\interp{E(C)}_{\textup{v}}=\interp{C}_{\textup{c}}$ for any $C\in\cat{CQC}\vert_{\mathcal{G}}(n,n)$.
\end{lemma}
\begin{proof}
  By structural induction with
  \begin{multline*}
    \interp{E(C_1\diamond C_2)}_{\textup{v}}
    =\interp{E(C_1)\diamond E(C_2)}_{\textup{v}}
    =\interp{E(C_1)}_{\textup{v}}\diamond \interp{E(C_2)}_{\textup{v}}\\
    \overset{\textup{IH}}{=}\interp{C_1}_{\textup{c}}\diamond \interp{C_2}_{\textup{c}}
    =\interp{C_1\diamond C_2}_{\textup{c}}
  \end{multline*}
  where $\diamond$ is either $\circ$ or $\otimes$, and where the base cases can be checked manually.\qed
\end{proof}

\begin{lemma}\label{lem:decodinglambda}
  $\cat{CQC}\vdash D(\lambda^n(\alpha))=\mu^n(\alpha)$ for any $n\in\N$ and $\alpha\in\R$.
\end{lemma}
\begin{proof}
  By induction on $n$.
  The base cases where $n=0$ or $n=1$ are proved as follows.
  \begin{gather*}
    D(\lambda^0(\alpha))=D(\!\tf{alpha}\!)=\tf{alpha}=\mu^0(\alpha) \\[0.5em]
    D(\lambda^1(\alpha))=D(\tf{zalpha})=\tf{calpha}=\mu^1(\alpha)
  \end{gather*}
  
  The induction case uses the fact that $\cat{CQC}\vdash D(\sigma_{1,1})=\sigma_{1,1}$ (see \cref{eq:swap}) and is proved as follows.
  \begin{gather*}
    D(\lambda^{n+2}(\alpha))
    =D\left(\tf{decoding_mu_01}\right) \\[0.5em]
    \overset{\textup{IH}}{=}\tf{decoding_mu_02} \\[0.5em]
    \eqoneref{eq:swap}\tf{decoding_mu_03}
    =\mu^{n+2}(\alpha)
  \end{gather*}\qed
\end{proof}

\begin{lemma}\label{lem:dec}
  If $\cat{QC}\vdash C_1=C_2$ then $\cat{CQC}\vdash D(C_1)=D(C_2)$.
\end{lemma}
\begin{proof}
  This is enough to prove the statement for each relation $C_1=C_2$ in $\et{R}_{\textup{v}}$.
  \cref{eq:2pibare,,eq:additionbare,,eq:hhbare,,eq:eulerbare} of $\et{R}_{\textup{v}}$ are directly decoded as \cref{eq:2pi,,eq:addition,,eq:hh,,eq:euler} in $\et{R}_{\textup{c}}$.
  \cref{eq:swapbare} is decoded as follows.
  \begin{gather*}
    D\left(\tf{decoding_swap_01}\right)
    =\tf{decoding_swap_02}
    \eqoneref{eq:swap}\tf{decoding_swap_03}
    =\tf{decoding_swap_04}\\[0.5em]
    \eqoneref{eq:swap}\tf{decoding_swap_05}
    =\tf{decoding_swap_06}\\[0.5em]
    \eqthreeref{eq:hh}{eq:2pi}{eq:addition}\tf{decoding_swap_07}\\[0.5em]
    =D\left(\tf{decoding_swap_08}\right)
    =D\left(\tf{decoding_swap_09}\right)
  \end{gather*}
  \cref{eq:0bare} is decoded as follows.
  \begin{gather*}
    D\left(\tf{decoding_c0_01}\right)
    =\tf{decoding_c0_02}
    \eqoneref{eq:0}\tf{decoding_c0_03}
    =\tf{decoding_c0_04}
    =D\left(\tf{decoding_c0_04}\right)
  \end{gather*}
  \cref{eq:cadditionbare} is decoded as follows.
  \begin{gather*}
    D\left(\tf{decoding_caddition_01}\right)
    =\tf{decoding_caddition_02}
    =\tf{decoding_caddition_03}
    \eqoneref{eq:addition}\tf{decoding_caddition_04}
    =D\left(\tf{decoding_caddition_05}\right)
  \end{gather*}
  \cref{eq:zcommutbare} is decoded as follows.
  \begin{gather*}
    D\left(\tf{decoding_zcommut_01}\right)
    =\tf{decoding_zcommut_02}
    \eqoneref{eq:hh}\tf{decoding_zcommut_03}\\[0.5em]
    =\tf{decoding_zcommut_04}
    \eqtworef{eq:addition}{eq:2pi}\tf{decoding_zcommut_05}
    \eqoneref{eq:hh}\tf{decoding_zcommut_06}
    =D\left(\tf{decoding_zcommut_07}\right)
  \end{gather*}
  \cref{eq:czbare} is decoded as follows.
  \begin{gather*}
    D\left(\tf{decoding_cz_01}\right)
    =\tf{decoding_cz_02}
    \eqoneref{eq:hh}\tf{decoding_cz_03} \\[0.7em]
    \eqcref{prop:ctrlunfolding}\tf{decoding_cz_04}
    \eqcref{lem:decodinglambda}D\left(\tf{decoding_cz_05}\right)
  \end{gather*}
  \cref{eq:mc2pibare} is decoded as follows.
  \begin{gather*}
    D\left(\tf{decoding_mc2pi_01}\right)
    \eqcref{lem:decodinglambda}\tf{decoding_mc2pi_02}
    \eqcref{prop:ctrlunfolding}\tf{decoding_mc2pi_03}\\[0.7em]
    \eqoneref{eq:2pi}\tf{decoding_mc2pi_04}
    =\tf{decoding_mc2pi_05}
    =D\left(\tf{decoding_mc2pi_05}\right)
  \end{gather*}\qed
\end{proof}

\begin{lemma}\label{lem:decenc}
  $\cat{CQC}\vdash D(E(C))=C$ for any $C\in\cat{CQC}\vert_{\mathcal{G}}(n,n)$.
\end{lemma}
\begin{proof}
  By structural induction with
  \begin{equation*}
    D(E(C_1\diamond C_2))=D(E(C_1)\diamond E(C_2))=D(E(C_1))\diamond D(E(C_2))\overset{\textup{IH}}{=}C_1\diamond C_2
  \end{equation*}
  where $\diamond$ is either $\circ$ or $\otimes$, and where the base cases are proved as follows.
  \begin{gather*}
    D(E(g))=D(g)=g \quad \text{when } \quad g\in\left\{\textup{id}_1,\tf{alpha},\tf{h}\right\}\\[0.5em]
    D\left(E\left(\tf{calphagen}\right)\right)=D\left(\tf{zalpha}\right)=\tf{calphagen}\\[0em]
    D\left(E\left(\tf{ccpigen}\right)\right)=D\left(\tf{hcnoth}\right)=\tf{hcnoth_step}\eqoneref{eq:hh}\tf{ccpi}
  \end{gather*}\qed
\end{proof}

\begin{proof}[\cref{th:cqccompleteness}]
  Let $C_1,C_2\in\cat{CQC}(n,n)$ such that $\interp{C_1}_{\textup{c}}=\interp{C_2}_{\textup{c}}$. By \cref{lem:cqcreducing}, there are some $C_1',C_2'\in\cat{CQC}\vert_{\mathcal{G}}(n,n)$ such that $\cat{CQC}\vdash C_i=C_i'$.
  By \Cref{lem:interp} and completeness of $\cat{QC}$ (see \cref{th:qccompleteness}) we have $\cat{QC}\vdash E(C_1')=E(C_2')$. Then, by \Cref{lem:dec}, $\cat{CQC}\vdash D(E(C_1'))=D(E(C_2'))$ and \Cref{lem:decenc} concludes the proof as follows.
  \begin{multline*}
    \cat{CQC}\vdash C_1
    \eqcref{lem:cqcreducing}\hspace{0.8em}C_1'
    \eqcref{lem:decenc}\hspace{0.8em}D(E(C_1'))\\
    \eqcref{lem:dec}\hspace{0.8em}D(E(C_2'))
    \eqcref{lem:decenc}\hspace{0.8em}C_2'
    \eqcref{lem:cqcreducing}\hspace{0.8em}C_2
  \end{multline*}\qed
\end{proof}

\begin{proposition}\label{prop:minimality}
  \cref{eq:2pi,eq:addition,eq:swap,eq:hh,eq:euler} of $\et{R}_{\textup{c}}$ are necessary.
\end{proposition}
\begin{proof}
  \cref{eq:2pi} is the only equation acting on 0 qubits that transforms a circuit containing some global phase gates into one that contains none.
  At least one instance of \cref{eq:addition} containing a parameter $\alpha$ is necessary for any $\alpha\in\R\backslash\{2\pi\}$, otherwise there is no equation acting on 0 qubits that transforms a circuit containing some \tf{alpha} gates into one that contains none.
  \cref{eq:hh} is the only equation acting on 1 qubit or fewer that transforms a circuit containing some \tf{h} gates into one that contains none.
  \cref{eq:swap} is the only equation that does not preserve the parity of \smalltf{swap} gates.
  \cref{eq:euler} is the only equation that does not preserve the parity of \tf{h} gates.\qed
\end{proof}


\section{Proof of \cref{prop:aqcpoints}}\label[appendix]{app:aqcpoints}

\begin{equation}\label{eq:notnot}
    \tf{notnot_01}=\tf{notnot_03}
\end{equation}

\begin{equation}\label{eq:winitccalpha}
    \tf{winitccalpha_01}=\tf{winitccalpha_04}
\end{equation}

\begin{equation}\label{eq:winitcccpi}
    \tf{winitcccpi_01}=\tf{winitcccpi_04}
\end{equation}

\begin{equation}\label{eq:winitch}
    \tf{winitch_01}=\tf{winitch_04}
\end{equation}

\begin{equation}\label{eq:binitcalpha}
    \tf{binitcalpha_01}=\tf{binitcalpha_07}
\end{equation}

\begin{equation}\label{eq:binitccpi}
    \tf{binitccpi_01}=\tf{binitccpi_10}
\end{equation}

\begin{equation}\label{eq:binitccalpha}
    \tf{binitccalpha_01}=\tf{binitccalpha_06}
\end{equation}

\begin{equation}\label{eq:binitcccpi}
    \tf{binitcccpi_01}=\tf{binitcccpi_04}
\end{equation}

\begin{equation}\label{eq:binitch}
    \tf{binitch_01}=\tf{binitch_04}
\end{equation}

\begin{lemma}\label{lem:aqcpointsbasecases}
    The following equations are derivable for any $C\in\cat{CQC}\vert_{\mathcal{G}}(n,n)$.
    \begin{equation*}
        \cat{AQC}\vdash\tf{winitcc}=\tf{winitidid}
        \hspace{5em}
        \cat{AQC}\vdash\tf{binitcc}=\tf{binitc}
    \end{equation*}
\end{lemma}
\begin{proof}
    By induction on $C$. The base cases are proved as \cref{eq:winitcalpha,eq:winitccalpha,eq:winitcccpi,eq:winitch,eq:binitcalpha,eq:binitccalpha,eq:binitcccpi,eq:binitch} at the end of the section. If $C=C_2\circ C_1$, we have the following derivations.
    \begin{equation*}
        \tf{cqcfalseseq_01}
        =\tf{cqcfalseseq_02}
        \overset{\textup{IH}}{=}\tf{cqcfalseseq_03}
        \overset{\textup{IH}}{=}\tf{cqcfalseseq_04}
    \end{equation*}
    \begin{equation*}
        \tf{cqctrueseq_01}
        =\tf{cqctrueseq_02}
        \overset{\textup{IH}}{=}\tf{cqctrueseq_03}
        \overset{\textup{IH}}{=}\tf{cqctrueseq_04}
        =\tf{cqctrueseq_05}
    \end{equation*}
    If $C=C_1\otimes C_2$, we have the following derivations.
    \begin{equation*}
        \tf{cqcfalsepar_01}
        =\tf{cqcfalsepar_02}
        =\tf{cqcfalsepar_03}
        \overset{\textup{IH}}{=}\tf{cqcfalsepar_04}
        \overset{\textup{IH}}{=}\tf{cqcfalsepar_05}
    \end{equation*}
    \begin{equation*}
        \tf{cqctruepar_01}
        =\tf{cqctruepar_02}
        =\tf{cqctruepar_03}
        \overset{\textup{IH}}{=}\tf{cqctruepar_04}
        \overset{\textup{IH}}{=}\tf{cqctruepar_05}
        =\tf{cqctruepar_06}
    \end{equation*}
    \qed
\end{proof}

\begin{proof}[\cref{prop:aqcpoints}]
    We need to prove the same equations as in \cref{lem:aqcpointsbasecases} but for any $C\in\cat{AQC}_\textup{clean}(n,n)$. To do so, first apply \cref{prop:pulloutancilla} to get a circuit $C'\in\cat{CQC}(n',n')$ and then \cref{lem:cqcreducing} to get a circuit $C''\in\cat{CQC}\vert_{\mathcal{G}}(n',n')$. Then, we have the following derivation for the white initialization.
    \begin{gather*}
        \tf{aqcfalse_01}
        \eqcref{prop:pulloutancilla}\tf{aqcfalse_02}
        \eqoneref{eq:conjugationancilla}\tf{aqcfalse_03} \\[0.4em]
        \eqcref{lem:cqcreducing}\tf{aqcfalse_04} 
        \eqcref{lem:aqcpointsbasecases}\tf{aqcfalse_05}
        \eqoneref{eq:winitwterm}\tf{aqcfalse_06}
    \end{gather*}
    
    And we have the following derivation for the black initialization.
    \begin{gather*}
        \tf{aqctrue_01}
        \eqcref{prop:pulloutancilla}\tf{aqctrue_02}
        \eqoneref{eq:conjugationancilla}\tf{aqctrue_03}
        \eqcref{lem:cqcreducing}\tf{aqctrue_04} \\[0.4em]
        \eqcref{lem:aqcpointsbasecases}\tf{aqctrue_05}
        \eqcref{lem:cqcreducing}\tf{aqctrue_06}
        \eqcref{prop:pulloutancilla}\tf{aqctrue_07}
    \end{gather*}
    \qed
\end{proof}

\begin{proof}[\cref{eq:notnot}]
    \begin{gather*}
        \tf{notnot_01}
        =\tf{notnot_02}
        \eqthreeref{eq:hh}{eq:addition}{eq:2pi}\tf{notnot_03}
    \end{gather*}\qed
\end{proof}

\begin{proof}[\cref{eq:winitccalpha}]
    \begin{gather*}
        \tf{winitccalpha_01}
        \eqoneref{eq:ccalphareduce}\tf{winitccalpha_02}\\[0.4em]
        \eqtworef{eq:winitcalpha}{eq:winitccpi}\tf{winitccalpha_03}
        \eqthreeref{eq:hh}{eq:addition}{eq:0}\tf{winitccalpha_04}
    \end{gather*}\qed
\end{proof}

\begin{proof}[\cref{eq:winitcccpi}]
    \begin{gather*}
        \tf{winitcccpi_01}
        \eqoneref{eq:cccpireduce}\tf{winitcccpi_02}\\[0.4em]
        \eqoneref{eq:winitccalpha}\tf{winitcccpi_03}
        \eqthreeref{eq:hh}{eq:addition}{eq:2pi}\tf{winitcccpi_04}
    \end{gather*}\qed
\end{proof}

\begin{proof}[\cref{eq:winitch}]
    \begin{gather*}
        \tf{winitch_01}
        \eqoneref{eq:chreduce}\tf{winitch_02}\\[0.4em]
        \eqoneref{eq:winitccpi}\tf{winitch_03}
        \eqthreeref{eq:hh}{eq:addition}{eq:0}\tf{winitch_04}
    \end{gather*}\qed
\end{proof}

\begin{proof}[\cref{eq:binitcalpha}]
    \begin{gather*}
        \tf{binitcalpha_01}
        =\tf{binitcalpha_02}
        =\tf{binitcalpha_03}
        \eqoneref{eq:xcalpha}\tf{binitcalpha_04}\\[0.4em]
        \eqoneref{eq:winitcalpha}\tf{binitcalpha_05}
        =\tf{binitcalpha_06}
        =\tf{binitcalpha_07}
    \end{gather*}\qed
\end{proof}

\begin{proof}[\cref{eq:binitccpi}]
    \begin{gather*}
        \tf{binitccpi_01}
        =\tf{binitccpi_02}
        \eqfourref{eq:hh}{eq:2pi}{eq:addition}{eq:notnot}\tf{binitccpi_03}\\[0.4em]
        =\tf{binitccpi_04}\\[0.4em]
        =\tf{binitccpi_05}\\[0.4em]
        \eqoneref{eq:ccalphareduce}\tf{binitccpi_06}
        \eqoneref{eq:2pi}\tf{binitccpi_07}
        \eqoneref{eq:hh}\tf{binitccpi_08}\\[0.4em]
        \eqoneref{eq:winitccpi}\tf{binitccpi_09}
        =\tf{binitccpi_10}
    \end{gather*}\qed
\end{proof}

\begin{proof}[\cref{eq:binitccalpha}]
    \begin{gather*}
        \tf{binitccalpha_01}
        \eqoneref{eq:ccalphareduce}\tf{binitccalpha_02}\\[0.4em]
        \eqoneref{eq:binitccpi}\tf{binitccalpha_03}
        \eqoneref{eq:xcalpha}\tf{binitccalpha_04}\\[0.4em]
        \eqtworef{eq:addition}{eq:0}\tf{binitccalpha_05}
        \eqthreeref{eq:hh}{eq:addition}{eq:2pi}\tf{binitccalpha_06}
    \end{gather*}\qed
\end{proof}

\begin{proof}[\cref{eq:binitcccpi}]
    \begin{gather*}
        \tf{binitcccpi_01}
        \eqoneref{eq:cccpireduce}\tf{binitcccpi_02}\\[0.4em]
        \eqoneref{eq:binitccalpha}\tf{binitcccpi_03}
        \eqoneref{eq:ccalphareduce}\tf{binitcccpi_04}
    \end{gather*}\qed
\end{proof}

\begin{proof}[\cref{eq:binitch}]
    \begin{gather*}
        \tf{binitch_01}
        \eqoneref{eq:chreduce}\tf{binitch_02}\\[0.4em]
        \eqoneref{eq:binitccpi}\tf{binitch_03}
        \eqoneref{eq:heulervar}\tf{binitch_04}
    \end{gather*}\qed
\end{proof}


\section{Variant of the cosine-sine decomposition}\label[appendix]{app:csdvariant}

We provide a self-contained proof of the \emph{generalized cosine-sine decomposition (GCSD)} (introduced for a similar purpose in \cite{extensions}), which is a generalization of the \emph{cosine-sine decomposition (CSD)} \cite{csdhistory}. All considered matrices are complex, and we write $A:n\times m$ when the matrix $A$ has $n$ rows and $m$ columns. 

First, recall some usual decompositions of an arbitrary square matrix $A:n\times n$. The \emph{RQ-decomposition (RQD)} states that there exist a unitary matrix $Q:n\times n$ and an upper-triangular matrix $R:n\times n$ with real non-negative diagonal elements such that $A=RQ$. The \emph{QL-decomposition (QLD)} states that there exist a unitary matrix $Q:n\times n$ and a lower-triangular matrix $L:n\times n$ with real non-negative diagonal elements such that $A=QL$. The \emph{singular value decomposition (SVD)} states that there exist two unitary matrices $U,V:n\times n$ and a real diagonal matrix $D=\diag(d_1,d_2,\dots,d_n)$ satisfying $d_1\ge d_2\ge\dots\ge d_n\ge 0$ and such that $A=UDV$ \cite{matrixalgebra}.

\begin{lemma}\label{lem:triangulardiagonal}
    If $L:n\times n$ is a lower-triangular matrix and $L^\dagger L$ is diagonal with non-zero diagonal elements, then $L$ is diagonal. If $R:n\times n$ is an upper-triangular matrix and $RR^\dagger$ is diagonal with non-zero diagonal elements, then $R$ is diagonal.
\end{lemma}
\begin{proof}
    We prove the statement for a lower-triangular matrix $L:n\times n$ by induction on $n$ (the statement for an upper-triangular matrix $R$ follows because $R^\dagger$ is a lower-triangular matrix such that $(R^\dagger)^\dagger R^\dagger=RR^\dagger$ is diagonal with non-zero elements). The base case $n=1$ is trivial. For the induction case where $L:(n+1)\times(n+1)$, let $D\defeq L^\dagger L$. Notice that $D_{n+1,n+1}=|L_{n+1,n+1}|^2\ne 0$ and $D_{i,n+1}=\overline{L_{n+1,i}}L_{n+1,n+1}=0$ for any $i\in\natset{n}$, which implies $L_{n+1,n+1}\ne 0$ and $L_{n+1,i}=0$. Hence, all non-diagonal elements of the last row and last column of $L$ are zero, and we can use the IH on the submatrix obtained by deleting the last row and the last column of $L$.
\end{proof}

\begin{lemma}[\cite{csdhistory}]\label{lem:csd}
    Given a unitary matrix $U:2n\times 2n$, the \emph{cosine-sine decomposition (CSD)} states that there exist unitary matrices $A_0,A_1,B_0,B_1:n\times n$, and real diagonal matrices $D_c,D_s:n\times n$ with non-negative elements such that $D_c^2+D_s^2=\idmat{n}$ and 
    \begin{gather*}
        U=\left[\begin{array}{c|c}
            A_0 & 0 \\
            \hline
            0 & A_1
        \end{array}\right]\cdot\left[\begin{array}{c|c}
            D_c & -D_s\\
            \hline
            D_s & D_c
        \end{array}\right]\cdot\left[\begin{array}{c|c}
            B_0 & 0 \\
            \hline
            0 & B_1
        \end{array}\right].
    \end{gather*}
\end{lemma}

\begin{lemma}\label{lem:gcsd}
    Given a unitary matrix $U:(2n-k)\times (2n-k)$ where $n\ge k\ge0$, the \emph{generalized cosine-sine decomposition (GCSD)} states that there exist unitary matrices $A_0,B_0:(n-k)\times (n-k)$, $A_1,B_1:n\times n$ and real diagonal matrices $D_c,D_s:(n-k)\times (n-k)$ with non-negative elements such that $D_c^2+D_s^2=\idmat{n-k}$ and
    \begin{gather*}
        \left[\begin{array}{c|c}
            \idmat{k} & 0 \\
            \hline
            0 & U
        \end{array}\right]=\left[\begin{array}{c|c|c}
            \idmat{k} & 0 & 0 \\
            \hline
            0 & A_0 & 0 \\
            \hline
            0 & 0 & A_1
        \end{array}\right]\cdot\left[\begin{array}{c|c|c|c}
            \idmat{k} & 0 & 0 & 0 \\
            \hline
            0 & D_c & 0 & -D_s\\
            \hline
            0 & 0 & \idmat{k} & 0\\
            \hline
            0 & D_s & 0 & D_c
        \end{array}\right]\cdot\left[\begin{array}{c|c|c}
            \idmat{k} & 0 & 0 \\
            \hline
            0 & B_0 & 0 \\
            \hline
            0 & 0 & B_1
        \end{array}\right].
    \end{gather*}
\end{lemma}
\begin{proof}
    Let $\hat{U}\defeq \idmat{k}\oplus U$. We can split $U$ into four blocks $U_{00}:(n-k)\times (n-k)$, $U_{01}:(n-k)\times n$, $U_{10}:n\times (n-k)$ and $U_{11}:n\times n$. Notice that the unitarity of $U$ implies the unitarity of $\hat{U}$.
    \begin{gather*}
        \hat{U}=\left[\begin{array}{c|c|c}
            \idmat{k} & 0 & 0 \\
            \hline
            0 & U_{00} & U_{01} \\
            \hline
            0 & U_{10} & U_{11}
        \end{array}\right]
    \end{gather*}
    Then, let $A_0DB_0$ be the SVD of $U_{00}$, which yields the following decomposition.
    \begin{gather*}
        \hat{U}=\left[\begin{array}{c|c|c}
            \idmat{k} & 0 & 0 \\
            \hline
            0 & A_0DB_0 & U_{01} \\
            \hline
            0 & U_{10} & U_{11}
        \end{array}\right]
        =\left[\begin{array}{c|c|c}
            \idmat{k} & 0 & 0 \\
            \hline
            0 & A_0 & 0 \\
            \hline
            0 & 0 & \idmat{n}
        \end{array}\right]\cdot\left[\begin{array}{c|c|c}
            \idmat{k} & 0 & 0 \\
            \hline
            0 & D & A_0^\dagger U_{01} \\
            \hline
            0 & U_{10}B_0^\dagger & U_{11}
        \end{array}\right]\cdot\left[\begin{array}{c|c|c}
            \idmat{k} & 0 & 0 \\
            \hline
            0 & B_0 & 0 \\
            \hline
            0 & 0 & \idmat{n}
        \end{array}\right]
    \end{gather*}
    Using respectively the RQD and QLD, we get
    \begin{gather*}
        \left[\begin{array}{c}0 \\ \hline A_0^\dagger U_{01}\end{array}\right]=RQ
        \hspace{3em}\textup{and}\hspace{3em}
        \left[\begin{array}{c|c}0 & U_{10}B_0^\dagger\end{array}\right]=A_1L
    \end{gather*}
    where $R:n\times n$ is upper-triangular, $L:n\times n$ is lower-triangular and $A_1,Q:n\times n$ are unitaries. Then, the middle matrix can be further decomposed as follows.
    \begin{gather*}
        \left[\begin{array}{c|c}
            \begin{array}{c|c}
                \idmat{k} & 0 \\
                \hline
                0 & D
            \end{array} & RQ \\
            \hline
            A_1L & U_{11}
        \end{array}\right]
        =\left[\begin{array}{c|c|c}
            \idmat{k} & 0 & 0 \\
            \hline
            0 & \idmat{n-k} & 0 \\
            \hline
            0 & 0 & A_1
        \end{array}\right]\cdot\left[\begin{array}{c|c}
            \begin{array}{c|c}
                \idmat{k} & 0 \\
                \hline
                0 & D
            \end{array} & R \\
            \hline
            L & A_1^\dagger U_{11}Q^\dagger
        \end{array}\right]\cdot\left[\begin{array}{c|c|c}
            \idmat{k} & 0 & 0 \\
            \hline
            0 & \idmat{n-k} & 0 \\
            \hline
            0 & 0 & Q
        \end{array}\right]
    \end{gather*}
    The unitarity of $\hat{U}$ forces the elements of the diagonal matrix $D$ to be between $0$ and $1$, and the SVD sorts them from largest to smallest. Let $\ell\ge k$ be such that the first $\ell-k$ diagonal elements of $D$ are $1$s and let $D_{00}:(n-\ell)\times(n-\ell)$ be the submatrix of $D$ which has only elements strictly smaller than $1$, which implies
    \begin{gather*}
        \left[\begin{array}{c|c}
            \idmat{k} & 0 \\
            \hline
            0 & D
        \end{array}\right]=\left[\begin{array}{c|c}
            \idmat{\ell} & 0 \\
            \hline
            0 & D_{00}
        \end{array}\right].
    \end{gather*}
    Then, we can split $V\defeq A_1^\dagger U_{11}Q^\dagger:n\times n$ into four blocks $V_{00}:\ell\times\ell$, $V_{01}:\ell\times(n-\ell)$, $V_{10}:(n-\ell)\times\ell$ and $V_{11}:(n-\ell)\times(n-\ell)$. Moreover, the orthonormality of the $\ell$ first rows and columns of $\hat{U}$ forces the middle matrix to have the form
    \begin{gather*}
        \left[\begin{array}{c|c|c|c}
            \idmat{\ell} & 0 & 0 & 0 \\
            \hline
            0 & D_{00} & 0 & D_{01}\\
            \hline
            0 & 0 & V_{00} & V_{01}\\
            \hline
            0 & D_{10} & V_{10} & V_{11}
        \end{array}\right]
    \end{gather*}
    where $D_{01}:(n-\ell)\times(n-\ell)$ is the resulting upper-triangular submatrix of $R$ and $D_{10}:(n-\ell)\times(n-\ell)$ is the resulting lower-triangular submatrix of $L$. The unitarity of $\hat{U}$ implies that 
    \begin{gather*}
        D_{00}D_{00}^\dagger+D_{01}D_{01}^\dagger=\idmat{n-\ell}
        \hspace{3em}\textup{and}\hspace{3em}
        D_{00}^\dagger D_{00}+D_{10}^\dagger D_{10}=\idmat{n-\ell}
    \end{gather*}
    where $D_{00}D_{00}^\dagger=D_{00}^\dagger D_{00}=D_{00}^2$ is diagonal with only non-negative strictly smaller than $1$ elements. Thus, $D_{01}D_{01}^\dagger$ and $D_{10}^\dagger D_{10}$ are diagonal with non-zero diagonal elements and \cref{lem:triangulardiagonal} forces $D_{01}$ and $D_{10}$ to be diagonal (with real non-negative diagonal elements as the diagonal elements of the RQD and QLD are so). Thus, $D_{00}^2+D_{01}^2=\idmat{n-\ell}=D_{00}^2+D_{10}^2$ which implies $D_s'\defeq D_{10}=D_{01}$. Notice that all diagonal elements of $D_s'$ are $>0$. Moreover, the unitarity of $\hat{U}$ implies $V_{01}D_s'=0$ and $D_s'V_{10}=0$, which forces $V_{01}=0$ and $V_{10}=0$.
    Let $D_c'\defeq D_{00}$. Again, the unitarity of $\hat{U}$ implies $D_s'D_c'+V_{11}D_s'=0$, which forces $V_{11}=-D_c'$. Then, the middle matrix can be further decomposed as follows.
    \begin{gather*}
        \left[\begin{array}{c|c|c|c}
            \idmat{\ell} & 0 & 0 & 0 \\
            \hline
            0 & D_c' & 0 & D_s'\\
            \hline
            0 & 0 & V_{00} & 0\\
            \hline
            0 & D_s' & 0 & -D_c'
        \end{array}\right]
        =\left[\begin{array}{c|c|c|c}
            \idmat{\ell} & 0 & 0 & 0 \\
            \hline
            0 & D_c' & 0 & -D_s'\\
            \hline
            0 & 0 & \idmat{\ell} & 0\\
            \hline
            0 & D_s' & 0 & D_c'
        \end{array}\right]\cdot\left[\begin{array}{c|c|c|c}
            \idmat{\ell} & 0 & 0 & 0 \\
            \hline
            0 & \idmat{n-\ell} & 0 & 0\\
            \hline
            0 & 0 & V_{00} & 0\\
            \hline
            0 & 0 & 0 & -\idmat{n-\ell}
        \end{array}\right]
    \end{gather*}
    Finally, we can pad $D_c'$ with $1$s and $D_s'$ with $0$s to get $D_c,D_s:(n-k)\times(n-k)$ such that
    \begin{gather*}
        \left[\begin{array}{c|c}
            \idmat{\ell} & 0 \\
            \hline
            0 & D_c'
        \end{array}\right]=\left[\begin{array}{c|c}
            \idmat{k} & 0 \\
            \hline
            0 & D_c
        \end{array}\right]
        \hspace{3em}\textup{and}\hspace{3em}
        \left[\begin{array}{c|c}
            0_\ell & 0 \\
            \hline
            0 & D_s'
        \end{array}\right]=\left[\begin{array}{c|c}
            0_k & 0 \\
            \hline
            0 & D_s
        \end{array}\right]
    \end{gather*}
    where $0_k$ represents the $k\times k$ zero matrix. All put together we get the following decomposition.
    \begin{gather*}
        \hat{U}=\left[\begin{array}{c|c|c}
            \idmat{k} & 0 & 0 \\
            \hline
            0 & A_0 & 0 \\
            \hline
            0 & 0 & A_1
        \end{array}\right]\cdot\left[\begin{array}{c|c|c|c}
            \idmat{k} & 0 & 0 & 0 \\
            \hline
            0 & D_c & 0 & -D_s\\
            \hline
            0 & 0 & \idmat{k} & 0\\
            \hline
            0 & D_s & 0 & D_c
        \end{array}\right]\cdot\left[\begin{array}{c|c|c}
            \idmat{n} & 0 & 0 \\
            \hline
            0 & V_{00} & 0\\
            \hline
            0 & 0 & -\idmat{n-\ell}
        \end{array}\right]\cdot\left[\begin{array}{c|c|c}
            \idmat{k} & 0 & 0 \\
            \hline
            0 & B_0 & 0 \\
            \hline
            0 & 0 & Q
        \end{array}\right]
    \end{gather*}
    And we are done with $B_1\defeq\left[\begin{array}{c|c}
            V_{00} & 0 \\
            \hline
            0 & -\idmat{n-\ell}
        \end{array}\right]Q$.\qed
\end{proof}


\section{Proof of \cref{prop:isoidentity}}\label[appendix]{app:isoidentity}

In this section we use the notation $\bigtf{wcontrol}\defeq\bigtf{wcontroldef}$ to depict a negative control. Notice that \cref{prop:aqcpoints} (together with \cref{eq:notnot}) implies that \tf{binit} and \tf{winit} are points of the white control.
\begin{equation*}
    \cat{AQC}\vdash\tf{binitwcc}=\tf{binitidid}
    \hspace{5em}
    \cat{AQC}\vdash\tf{winitwcc}=\tf{winitc}
\end{equation*}
Moreover, the notation \bigtf{bwcontrol} may represent either a black control or a white control.

\begin{proof}[\cref{prop:isoidentity}]
    By induction on $n$. If $n=0$, there is no \tf{winit} and we can directly use the completeness of $\cat{CQC}$. If $n=1$, then we can use the completeness of $\cat{CQC}$ to transform $C$ into a circuit $\ctrl(C')$ and \cref{prop:aqcpoints} proves the equation. Now, suppose $C\in\cat{CQC}(1+n+k,1+n+k)$. Let $K\defeq 2^k$ and $N\defeq 2^{n+k}$. As the equation is sound with respect to $\interp{\cdot}_{\textup{i}}$, there exists a unitary matrix $U:(2N-K)\times(2N-K)$ satisfying the following equation.
    \begin{equation*}
        \interp{C}_{\textup{i}}=\left[\begin{array}{c|c}
            I_{K} & 0 \\
            \hline
            0 & U
        \end{array}\right]
    \end{equation*}
    
    \cref{lem:gcsd} gives the following decomposition
    \begin{equation*}
        \interp{C}_{\textup{i}}=\left[\begin{array}{c|c|c}
            I_K & 0 & 0 \\
            \hline
            0 & A_0 & 0 \\
            \hline
            0 & 0 & A_1
        \end{array}\right]\cdot\left[\begin{array}{c|c|c|c}
            I_K & 0 & 0 & 0 \\
            \hline
            0 & D_c & 0 & -D_s\\
            \hline
            0 & 0 & I_K & 0\\
            \hline
            0 & D_s & 0 & D_c
        \end{array}\right]\cdot\left[\begin{array}{c|c|c}
            I_K & 0 & 0 \\
            \hline
            0 & B_0 & 0 \\
            \hline
            0 & 0 & B_1
        \end{array}\right]
    \end{equation*}
    where $A_0,B_0:(N-K)\times (N-K)$, $A_1,B_1:N\times N$ are unitary matrices and $D_c,D_s:(N-K)\times (N-K)$ are real diagonal matrices with non-negative elements satisfying $D_c^2+D_s^2=I_{N-K}$. Notice that there exists $\theta_i\in[0,\nicefrac{\pi}{2}]$ such that
    \begin{equation*}
        D_c=\textup{diag}(\cos(\theta_1),\dots,\cos(\theta_{N-K}))
        \hspace{1.5em}\text{and}\hspace{1.5em}
        D_s=\textup{diag}(\sin(\theta_1),\dots,\sin(\theta_{N-K})).
    \end{equation*}

    Let $\theta\in\R$ and $R_y(\theta)\in\cat{CQC}(1,1)$ be a circuit such that 
    \begin{equation*}
        \interp{R_y(\theta)}_{\textup{i}} = \left[\begin{array}{c|c}
            \cos(\theta) & -\sin(\theta) \\
            \hline
            \sin(\theta) & \cos(\theta)
        \end{array}\right]
    \end{equation*}
    then, the following decomposition holds.
    \begin{equation*}
        \left[\begin{array}{c|c|c|c}
            I_{K} & 0 & 0 & 0 \\
            \hline
            0 & \;D_c\; & 0 & -D_s\\
            \hline
            0 & 0 & I_{K} & 0\\
            \hline
            0 & D_s & 0 & D_c
        \end{array}\right]
        =\prod_{2^k-1<i<2^{n+k}}\ctrl^{b(i)}R_y(\theta_{i-2^k+1})
    \end{equation*}
    where $b(i)$ is the binary representation of $i$ and where $\ctrl^{b(i)}R_y(\theta)\in\cat{CQC}(1+n+k,1+n+k)$ is a controlled $R_y(\theta)$ whose target is the first qubit and whose $(1+j)$-th qubit is positively controlled if the $j$-th digit of $b(i)$ is $1$, and negatively controlled if the $j$-th digit of $b(i)$ is $0$.

    Additionally, let $C_{A_0},C_{B_0},C_{A_1},C_{B_1}\in\cat{CQC}(n+k,n+k)$ be such that
    \begin{gather*}
        \interp{C_{A_0}}_{\textup{i}}=\left[\begin{array}{c|c}
            I_K & 0 \\
            \hline
            0 & A_0
        \end{array}\right] \hspace{2em}
        \interp{C_{B_0}}_{\textup{i}}=\left[\begin{array}{c|c}
            I_K & 0 \\
            \hline
            0 & B_0
        \end{array}\right] \hspace{2em}
        \interp{C_{A_1}}_{\textup{i}}=A_1 \hspace{2em}
        \interp{C_{B_1}}_{\textup{i}}=B_1
    \end{gather*}
    
    Then, by completeness of $\cat{CQC}$, the following equation is derivable.
    \begin{gather*}
        \tf{csd_01}
        =\tf{csd_02}
    \end{gather*}

    The first two controlled circuits can be respectively removed and fired using \cref{prop:aqcpoints} (together with \cref{eq:notnot}). Then, we can use the induction hypothesis to remove $C_{B_0}$. Moreover, notice that every bitstring $b(i)$ satisfying $2^k-1<i<2^{n+k}$ has at least one $1$ in its first $n$ digits. This implies that each $\ctrl^{b(i)}R_y(\theta_{i-2^k+1})$ has at least one positive control on a wire starting with a \tf{winit} gate, and we can thus remove all of them using \cref{prop:aqcpoints} (together with \cref{eq:notnot}). Finally, the last two controlled circuits can be removed as the first two, which concludes the proof.\qed
\end{proof}


\section{Control functors for symmetric monoidal categories}\label[appendix]{app:controlsmc}

Throughout the paper we focused on symmetric monoidal categories (SMC) that are PROPs. Nevertheless, the notion of control functor and its diagrammatic representation extends straightforwardly to SMCs. The main difference is that a control functor is specific to a certain object. We generalize \cref{def:controlfunctor} as follows.

\begin{definition}[control functors]\label{def:controlfunctorsmc}
    Given an SMC $\cat{C}$, a \emph{control structure} is a family of control functors $\ctrl^{A}:\cat{C}_\textup{endo}\to\cat{C}_\textup{endo}$ parametrized by objects $A\in\mathcal{O}(\cat{C})$, mapping any object $B\in\mathcal{O}(\cat{C})$ to the object $A\otimes B$ and any morphism $f:B\to B$ to a morphism $\ctrl^A (f): A\otimes B\to A\otimes B$ while satisfying the coherence laws depicted in \cref{fig:coherencecsmc}.
\end{definition}

Notice that \cref{eq:controlidsmc,eq:controltensorsmc} guarantee that the control functor preserves the monoidal structure on objects, while \cref{eq:strenghtsmc,,eq:controlswapsmc,,eq:swapconjugationsmc} respectively correspond to \cref{eq:strenght,,eq:controlswap,,eq:swapconjugation}. Moreover, instantiating this generalized definition of control functor in the context of PROPs, we recover \cref{def:controlfunctor}. In particular, $\ctrl^n(C)$ represents $n$ nested applications of $\ctrl(\cdot)$ to $C$.

The notion of points can also be extended to arbitrary SMCs. Moreover, if $\ctrl^A$ and $\ctrl^B$ have points (respectively denoted $\tf{winit}_A,\tf{binit}_A$ and $\tf{winit}_B,\tf{binit}_B$), then $\tf{winit}_A\otimes\tf{ginit}_B$ and $\tf{binit}_A\otimes\tf{binit}_B$ are points for $\ctrl^{A\otimes B}$ for any $\tf{ginit}_B:I\to B$, as well as $\tf{ginit}_A\otimes\tf{winit}_B$ and $\tf{binit}_A\otimes\tf{binit}_B$ for any $\tf{ginit}_A:I\to A$.

\begin{figure*}[t]
    \fbox{\begin{minipage}{0.985\textwidth}\begin{center}
        \vspace{-0.3em}
        \hspace{-0.5em}\begin{subfigure}{0.39\textwidth}
            \begin{equation}\label{eq:controlidsmc}
                \begin{array}{ccc}
                    \scalebox{0.90}{$\ctrl^I(f)$}&=&\scalebox{0.90}{$f$}\\[.4em]
                    \tf{controlid_smc_left}&=&\tf{controlid_smc_right}
                \end{array}
            \end{equation}
        \end{subfigure}\hspace{3em}
        \begin{subfigure}{0.48\textwidth}
            \begin{equation}\label{eq:controltensorsmc}
                \begin{array}{ccc}
                    \scalebox{0.90}{$\ctrl^{A\otimes B}(f)$}&=&\scalebox{0.90}{$\ctrl^A(\ctrl^B (f))$}\\[.4em]
                    \tf{controltensor_smc_left}&=&\tf{controltensor_smc_right}
                \end{array}
            \end{equation}
        \end{subfigure}

        \hspace{-0.5em}
        \begin{subfigure}{0.41\textwidth}
            \begin{equation}\label{eq:strenghtsmc}
                \begin{array}{ccc}
                    \scalebox{0.90}{$\ctrl^A(f\otimes \textup{id}_B)$}&=&\scalebox{0.90}{$\ctrl^A(f)\otimes \textup{id}_B$}\\[.4em]
                    \tf{strenght_smc_left}&=&\tf{strenght_smc_right}
                \end{array}
            \end{equation}
        \end{subfigure}

        \hspace{-0.5em}
        \begin{subfigure}{0.65\textwidth}
            \begin{equation}\label{eq:controlswapsmc}
                \begin{array}{ccc}
                    \scalebox{0.90}{$\ctrl^{A\otimes B}(f)\circ(\sigma_{B,A}\otimes \textup{id}_{C})$}&=&\scalebox{0.90}{$(\sigma_{B,A}\otimes \textup{id}_{C})\circ\ctrl^{B\otimes A}(f)$}\\[.4em]
                    \tf{ctrlswap_smc_left}&=&\tf{ctrlswap_smc_right}
                \end{array}
            \end{equation}
        \end{subfigure}

        \begin{equation*}
            \scalebox{0.80}{$\ctrl^A((\textup{id}_{B}\otimes \sigma_{C,D}\otimes \textup{id}_{E})\circ f\circ(\textup{id}_{B}\otimes \sigma_{D,C}\otimes \textup{id}_{E}))=(\textup{id}_{A\otimes B}\otimes \sigma_{C,D}\otimes \textup{id}_{E})\circ \ctrl^A(f)\circ(\textup{id}_{A\otimes B}\otimes \sigma_{D,C}\otimes \textup{id}_{E})$}
        \end{equation*}
        \begin{equation}\label{eq:swapconjugationsmc}
            \hspace{1em}\tf{swapconjugation_smc_right}\hspace{3em}=\hspace{3em}\tf{swapconjugation_smc_left}
        \end{equation}
    \vspace{-0.2em}
    \end{center}\end{minipage}}
    \caption{\label{fig:coherencecsmc}Coherence laws of control functors for arbitrary SMCs. \cref{eq:controlidsmc,eq:controltensorsmc,eq:strenghtsmc,eq:controlswapsmc} are defined for any $f\in\cat{C}(C,C)$. \cref{eq:swapconjugationsmc} is defined for any $f\in\cat{C}(B\otimes C\otimes D\otimes E,B\otimes C\otimes D\otimes E)$.}
\end{figure*}

\end{document}